\def\ispreprint{1}  

\if \ispreprint1
\documentclass[preprint,authoryear,5p]{elsarticle}  
\else
\documentclass[preprint,authoryear,12pt]{elsarticle}  
\fi

\usepackage[utf8]{inputenc}
\usepackage{lineno} 
\usepackage{amssymb}
\usepackage{amsmath,bm}
\usepackage{graphicx}
\usepackage{hyperref} 
\usepackage{setspace}
\usepackage{subcaption}
\usepackage{appendix}
\usepackage{multicol}
\usepackage{multirow}

\begin{document}

\title{Surface particle motions excited by a low velocity normal impact into a granular medium
}
\author[a1]{Max Neiderbach\corref{cor1}}
\ead{mneiderb@u.rochester.edu}
\author[a1]{Bingcheng Suo}
\ead{bsuo@u.rochester.edu}
\author[a1]{Esteban Wright}
\ead{ewrig15@ur.rochester.edu}
\author[a1]{A. C. Quillen}
\ead{alice.quillen@rochester.edu}
\author[a2]{Mokin Lee}
\ead{mlee80@u.rochester.edu}
\author[a2]{Peter Miklavcic}
\ead{pmiklavc@ur.rochester.edu}
\author[a2]{Hesam Askari}
\ead{askari@rochester.edu}
\author[a3]{Paul S\'anchez}
\ead{diego.sanchez-lana@colorado.edu}

\address[a1]{Department of Physics and Astronomy, University of Rochester, Rochester, NY 14627, USA}

\address[a2]{Department of Mechanical Engineering, University of Rochester, Rochester, NY 14627, USA}


\address[a3]{Colorado Center for Astrodynamics Research, The University of Colorado Boulder, 3775 Discovery Drive, 429 UCB - CCAR, Boulder, CO 80303, USA}

\cortext[cor1]{Corresponding author}


\begin{abstract}
In laboratory experiments, high speed videos are used to detect and track mm-size surface particle motions caused by a low velocity normal impact into sand. Outside the final crater radius and prior to the landing of the ejecta curtain,  particle displacements are measured via particle tracking  velocimetry and with a cross-correlation method. Surface particles rebound and are also permanently displaced with both peak and permanent displacements rapidly decaying as a function of distance from the crater center. The surface begins to move before most of the ejecta curtain has landed, but continues to move after the subsurface seismic pulse has decayed. Ray angles for surface and subsurface velocities are similar to those described by a Maxwell's Z-model. This implies that the flow field outside the crater excavation region is a continuation of the crater excavation flow. The ratio of final particle displacement to crater radius resembles that measured for other impact craters.

\end{abstract}

\maketitle

\if\ispreprint1
\else 
\linenumbers 
\fi

\section{Introduction}

Apollo-class Near-Earth Asteroid binary (65803) Didymos is the target of the international collaboration Asteroid Impact \& Deflection Assessment, also known as AIDA, that supports the development and data interpretation of NASA's Double Asteroid Redirection Test (DART) mission \citep{Cheng_2018_DART,Rivkin_2021,Cheng_2020} and the European Space Agency's Hera mission \citep{Michel_2022}.
The goals of these missions to the potentially hazardous binary asteroid Didymos include measuring the momentum transfer efficiency and the resulting deflection from a hyper-velocity impact. 
DART will be the first high-speed impact experiment on an asteroid at a scale capable of changing the path of a natural celestial body.  Imaging during the impact, which will occur in September 2022, will be carried out by the accompanying 6U CubeSat Light Italian CubeSat for Imaging of Asteroids (LICIACube; \citealt{Dotto_2021}). 
On asteroids, 
impact-generated seismic motions cause crater erasure and degradation (e.g.,  \citealt{thomas05}) and move boulders near the site of impact \citep{Arakawa_2020,Honda_2021}.   
The LICIACube mission presents an opportunity to directly observe these processes taking place. 

On April 5th 2019, Hayabusa2's Small Carry-on Impactor (SCI) carried out an artificial impact
at an impact velocity of 2 km/s onto Near-Earth object 162173 Ryugu \citep{Arakawa_2020}. 
The SCI impact formed an impact crater with a 
rim radius of $R_{rim}= 8.8$ m \citep{Arakawa_2020,Honda_2021}.
Comparison of pre- and post-impact images revealed boulder displacements of a few centimeters in the region between 15 and 30 m from the impact center \citep{Honda_2021}.

Impact crater properties such as volume and radius can be described with power law relations that depend on dimensionless parameters such as impact velocity, gravity and substrate material strength 
\citep{Melosh_1985,Schmidt_1987,holsapple93,housen11}.  
The `point source assumption', as described by \citet{housen11}, is that important phenomena occur at radii greater than the impactor radius during crater formation.   This is connected to the concept of `late-stage equivalence’  \citep{Schmidt_1987}  where the assumption is that phenomena relevant for estimating crater size and the ejecta distribution occurs later in the crater formation process when the crater is widening, or equivalently at radii greater than the impactor radius.  Both of these concepts center on the proposal that a single parameter $C$, dependent on a product of powers of the impactor velocity, density and radius, governs the cratering process. The parameter $C$ is then used to develop scaling laws for crater and ejecta properties \citep{Schmidt_1987,housen11}. The assumption of `late-stage equivalence’ may also be appropriate for low velocity impacts in microgravity \citep{Celik_2022}.

If the properties of the crater are independent of substrate material strength, the crater is said to be in a gravity-dominated scaling regime (e.g., see section 2 by \cite{Celik_2022}). 
Because dry sand is cohesionless, impact craters into sand follow gravity scaling at all event scales \citep{holsapple93,housen11}. 
Recent discrete-element method (DEM) simulations \citep{Celik_2022} find that crater scaling relations in the gravity regime extend to low velocity impacts into granular media at microgravity. Thus low velocity laboratory impact experiments into a granular system, such as sand, could be relevant for understanding high velocity impact craters in low-g granular systems, including those on rubble asteroids (e.g., \cite{Wright_2020b,Cheng_2020,Wright_2022,Quillen_2022}).   Laboratory impacts with kinetic energy of about 1000 J may be relevant for understanding the DART mission impact that will have kinetic energy that is about 10 million times larger. 
Despite the extreme difference in size, materials, crater excavation processes, and rim uplift (e.g., \cite{Roddy_1978}), large km-sized craters in the gravity scaling regime may exhibit similarities to similar impact processes for smaller craters but in laboratory sand or on rubble asteroids.

High velocity impact craters can be divided into three sequential stages \citep{Melosh_1985, Melosh_1999}. In the first stage, the impact generates a shock which propagates outward from the site of impact and material is compressed.  
In the second stage a transient crater is excavated.  
Lastly the crater is modified through relaxation processes such as slumping and erosion.  
The contact and compression phase is the shortest and during this phase,  material in the target moves outward along directions that are approximately radial from the site of impact.  
Low velocity laboratory impacts into millet exhibit an analogous phase with a seismic subsurface pulse rapidly propagating radially away from the site of impact \citep{Quillen_2022}. 
In contrast with the long time scales involved in crater excavation, energy and momentum are rapidly transferred via shockwaves to the target itself \citep{Melosh_1999}.
Laboratory measurements of embedded accelerometers in granular systems support a similar interpretation as they measure subsurface seismic pulses for both high and low velocity impact experiments \citep{yasui15,Matsue_2020,Quillen_2022}. 

The compressive shock at first propagates outward radially but then upon the arrival of an expansion wave 
that is affected by the free surface, the substrate 
particle velocity is significantly reduced and the direction of the velocity vectors of the shocked materials is changed \citep{Melosh_1985,Kurosawa_2019}. 
Pressure gradients behind the shock deflect the particle trajectories toward the surface, resulting in curved trajectories.  
Maxwell's Z-model \citep{Maxwell_1977,Croft_1980,Kurosawa_2019} is a simple incompressible power-law flow model that captures many properties of the flow associated with crater excavation \citep{Austin_1981} including crater excavation in laboratory impact experiments into sand \citep{Yamamoto_2009}.

During the transient growth phase of the crater, underlying materials outside the crater are deformed. The crater rim is raised compared to the original surface level, not only because of the ejecta blanket but because there is deformation and structural uplift in the underlying material \citep{Roddy_1978,Croft_1980,Melosh_1985,Sturm_2016}.  If rocky materials are stratified horizontally prior to the impact, post impact strata are uplifted in the crater rim \citep{Roddy_1978}.
Crater rim uplift is due to injected material and dilation of the excavation cavity wall \citep{Roddy_1978,Croft_1980,melosh89,Poelchau_2009,Sharpton_2014,Sturm_2016}. 

By scaling physical parameters at the crater radius, \citet{Quillen_2022} estimated the peak displacement in the subsurface seismic pulse in their low velocity normal impact experiments into sand.  The displacement amplitude was on the order of a few millimeters, presenting a challenge for direct measurement if similar movement is observed on the surface. 
In addition to propagation of a short duration seismic pulse, the surface motions outside the crater could also be affected by longer duration crater excavation processes, including crater uplift and expansion of the transient crater (e.g., \citet{Melosh_1985}). 
Recent simulations of low velocity oblique impact experiments into granular media have found that the dynamics near the surface, in a region denoted a `skin zone', differed from that below the surface \citep{Miklavcic_2022}.  Thus the connection between the short duration subsurface seismic pulse and longer timescale crater excavation processes may be complex. 


In this study we focus on detecting small surface motions outside the crater as a function of time in our laboratory impact experiments. Our goals are to better understand processes directly affecting the surface and connect them to motions beneath the surface. 
Our study is motivated by the recent detection of boulder movements on Ryugu by \citet{Honda_2021} and the forthcoming opportunity, represented by the DART impact, to see an impact take place on an asteroid with imaging by LICIACube.

In section \ref{sec:PTV}, we describe  detection of surface particles motions  excited via a low velocity normal laboratory impact into sand that are seen in high speed videos. In section \ref{sec:surface}, particle motions on the surface are compared to subsurface motions measured with embedded accelerometers.  We describe both subsurface and surface flow fields.  In section \ref{sec:discussion} we compare final particle displacements on the surface near the crater radius to those of other impact craters.  
 
\section{Experiments}
\label{sec:exp}

\begin{figure*}[ht]
\centering
\includegraphics[scale=0.5, trim = 0 0 0 0]{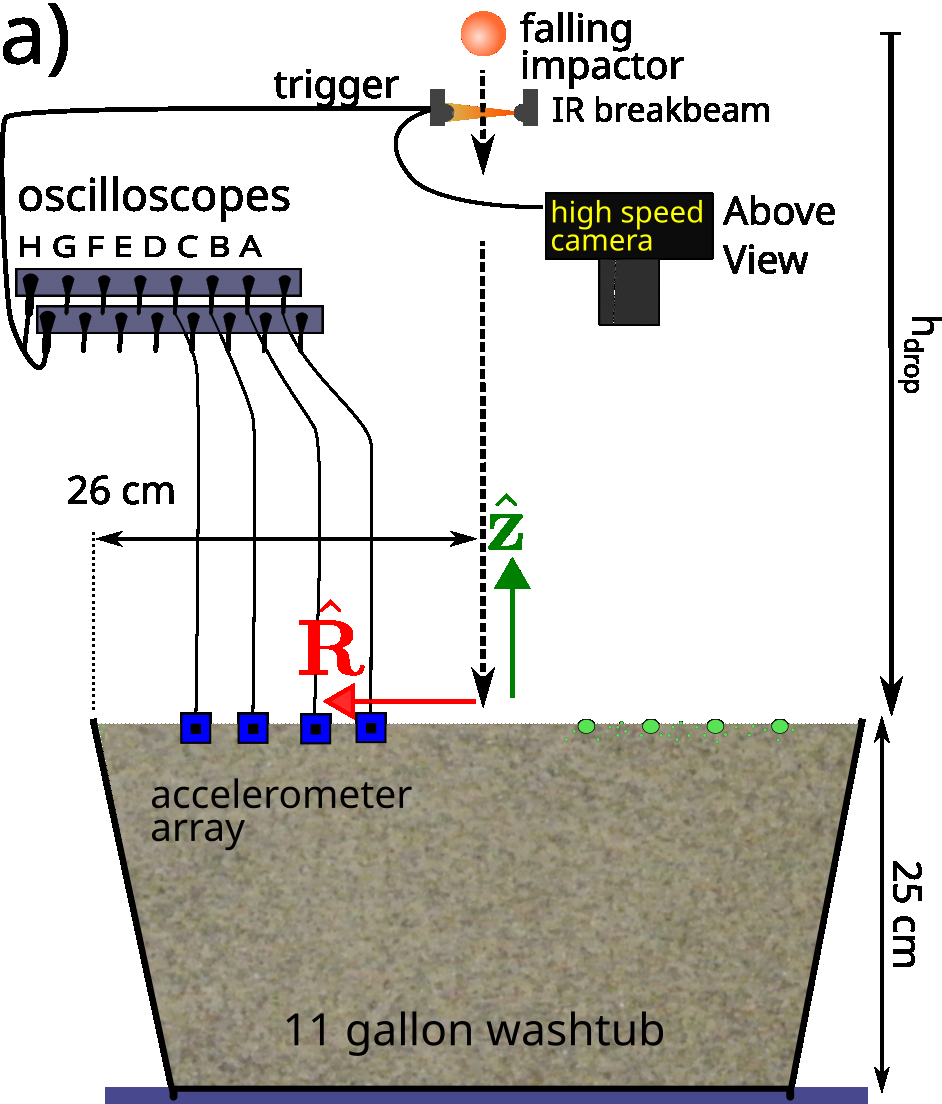}
\includegraphics[scale=0.55,trim= 0 -30 0 0 ,clip]{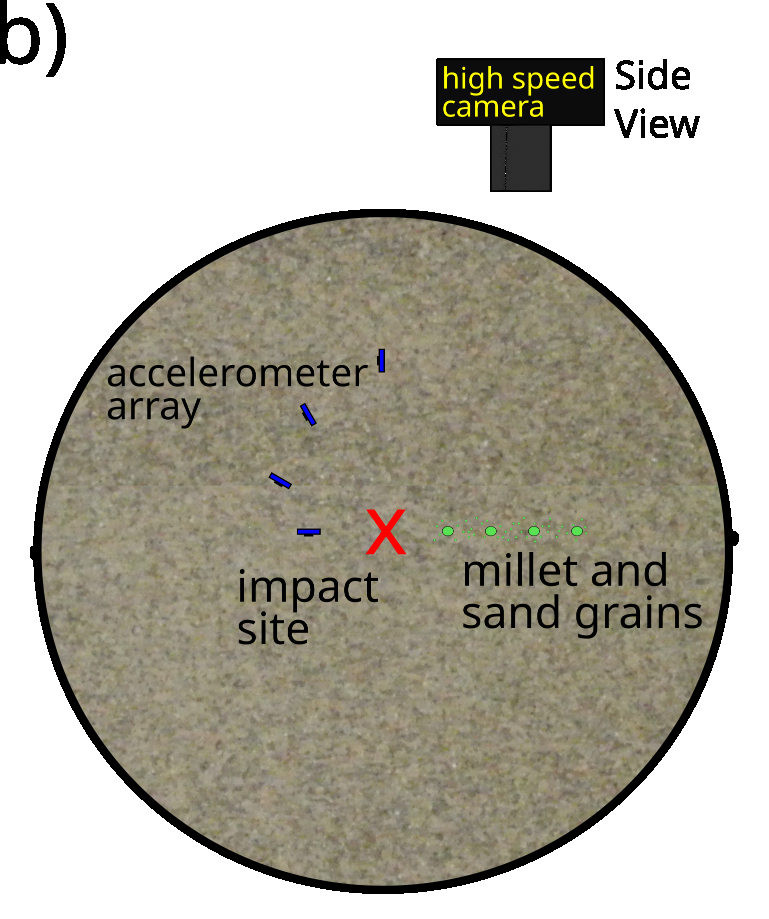}
\caption{a) Side view illustration of experiment. b) Top view illustration of experiment. Accelerometers are positioned at varying distances from impact.  Here we show 4 accelerometers that are resting on the surface and at different radii. Some of our experiments have accelerometers at different depths. 
Individual green sand and millet grains are used to track the surface motion.  The green grains are resting on the surface and shown with their size enlarged.
The projectile is released from an electromagnet and passes through an IR break-beam sensor, triggering the accelerometers and high speed cameras.
\label{fig:exp_setup}
}
\end{figure*}

We perform experiments of low velocity normal impacts into sand and we measure the motion of grains on the surface.  Our experiments are similar to the low velocity normal impact experiments into sand and millet described by \citet{Quillen_2022} except we focus on videos taken of the surface with high speed cameras and use a higher impact velocity.   Illustrations of our experiment are shown in Figure \ref{fig:exp_setup}.

\subsection{Granular substrate}

We use a galvanized 41.6 liter (11 gallon) washtub with a diameter of 50.2 cm and depth of 25 cm to hold our granular material.
The tub is filled with sand with a bulk density of $\rho_s = 1.5$\ g cm$^{-3}$.  
We use sand because the small particle size facilitates accurately measuring particle displacements with cross correlation methods. 
The grain semi-major axis mean value is 0.3 mm. 
The sand in the tub is raked and leveled prior to every impact experiment to remove compaction caused by previous impacts.
Rake tongs are 20 cm long and 4 cm apart (see \citet{Wright_2022} for further details on bed preparation).

\subsection{Impactor, drop height, crater radius and depth}

Our impactor is a glass marble with mass $M_g$ of $86.5$ g and radius $R_p$ of $19.9$ mm.
The density of our impactor is $2.58$ g/cm$^3$, which results in substrate to impactor density ratio of $\pi_4 = \rho_s/\rho_p = 0.58$. 
The impactor is initially suspended above the granular material with an electromagnet at a drop height $h_{\rm drop} = $ 176 cm.
Our impactor is accelerated by gravity alone once the electromagnet is switched off.

In all of our experiments the impacts were normal to the surface and have the same drop height giving an impact velocity of $v_{\rm imp} \approx \sqrt{2gh_{\rm drop}} \approx 5.88$ m/s.
Impacts occur at the center of the washtub.

The crater radius was found by measuring the diameter of the crater rim with a ruler. 
We found the crater radii were consistently $R_c \sim$5.5 cm for our experiments.
This value agreed with those measured from still images taken from above the experiment post-impact and using \texttt{ImageJ}\footnote{\url{https://imagej.nih.gov/ij/}} to measure the diameter using the circle tool.

The crater depth is measured from the bottom of the crater to the top of the crater rim and is $h_c \sim$2.2 cm.  After removing the impactor, we measured the depth by placing a toothpick in the base of the crater and marking it with a level straightedge resting across the crater rim.  Removing the impactor disturbs the crater.  For our low velocity impacts, the ratio of projectile to crater radius $R_p/R_c \sim 0.36$ is large and removing impactor disturbs the crater.  Consequently $h_c$ is only roughly measured. 

Experiment parameters are summarized in Table \ref{tab:exp_constants}
where we also include dimensionless $\pi_2 \equiv g R_p/v_{imp}^2$ and $\pi_4 \equiv \rho_{s}/\rho_p $ scaling parameters commonly used in crater scaling relationships \citep{housen11,Celik_2022}.
We also list the time estimated for transient crater formation
$\tau_{ex} = \sqrt{R_c/g}$ \citep{Housen_1983,Melosh_1985}.

\begin{table}[ht]
    \centering
    \caption{Physical Values for experiments}
    \begin{tabular}{lcl}
    \hline
         Projectile mass & $M_p$ & 86.5 g\\
         Projectile radius & $R_p$ & 19.9 mm\\
         Projectile density & $\rho_p$ & 2.58 g cm$^{-3}$ \\
         Substrate density & $\rho_s$ & 1.5 g cm$^{-3}$  \\ 
         Washtub rim radius & $R_{\rm tub}$ & 25.1 cm\\
         Washtub depth & $H_{\rm tub}$ & 25 cm\\
         Drop height & $h_{\rm drop}$ & 176 cm\\
         Impact velocity & $v_{\rm imp} = \sqrt{2gh_{\rm drop}}$ & 588 cm/s \\
         Crater radius & $R_c$ & 5.5 cm \\
         Crater depth & $h_c$ & 2.2 cm \\
         $\pi_2$ & $g R_p/v_{\rm imp}^2$ & 3.32 $\times 10^{-2}$ \\
         $\pi_4$ & $\rho_{s}/\rho_p$ & 0.58  \\
         $\tau_{ex}$ & $\sqrt{R_c/g}$ & 75 ms \\
    \hline
    \end{tabular}
    \label{tab:exp_constants}
\end{table}

\subsection{Triggering of accelerometers and cameras}

The impactor passes through an IR break-beam sensor that is used to trigger the accelerometers and up to two Krontech Chronos 1.4 high speed cameras\footnote{\url{https://www.krontech.ca/chronos-1-4-resources}} which film the surface. 

We perform 3 different sets of analyses: measurements of acceleration from embedded accelerometers, tracking the displacement of individual millet and sand grains on the surface using high speed videos, and measuring displacement of the surface through cross correlation image analysis, also using high speed videos. 

We define $t=0$ in all our figures and throughout this paper as the time of impact.
The time of impact was determined from the frame the projectile touches the surface of the substrate in the high speed video. 
Since our impacts were highly repeatable between experiments, the time between triggering the cameras or accelerometers is consistent.
We use this consistency to set the impact time for experimental runs with only accelerometers and no high speed videos.

\subsection{High Speed Videos}

The high-speed videos are captured at a frame-rate of 4533 frames per second (fps). Video frames are $1280 \times 240$ pixels.
We take videos under two different types of lighting conditions.  
Videos used for cross correlation are lit with bright white lights and use a single camera to film the surface from above.
Videos used for particle tracking are lit with bright blue LEDs and uses two cameras.
One of the cameras is positioned above the washtub and the second camera is located at the side of the washtub at the level of the surface.
This viewpoint captures the vertical and radial motions of the granular surface. 

The particles we use for particle tracking velocimetry (PTV) analysis are painted with paints that fluoresce under the blue lights. 
We use painted millet grains (about 3 mm in diameter) and painted sand grains (about 0.5 mm in diameter), both of which are lightly embedded into the surface of the sand.
This is to prevent the millet grains from rolling during the impact. We restrict the rolling because our objective is to study the average surface motions at a specific radial distance rather than the variability introduced by grains tumbling. 

A list of the videos is given in Table \ref{tab:exp_videos_list}.  The video labels start with `PTV' if the videos are used for particle tracking and start with `CC'  if they are used for cross correlation.
The `a' or `s' that immediately follows the `PTV' indicates if the video was taken from the above-view or side-view camera, and the number indicates the experimental trial.
Side and above-view videos are in pairs because we used two cameras for videos used in PTV analysis. 
For the PTVs-compare experiment we simultaneously took a side-view video and accelerometer data. This experiment was used to compare tracked surface particle motions with accelerations measured with accelerometers. 
In Table \ref{tab:exp_videos_list} for each video or video sequence, we also list the Figures that use measurements from the video.


\begin{table*}[]
    \centering
    \caption{List of experiments using high-speed videos}
    \begin{tabular}{llllccc}
    \hline
        Videos & Viewpoint &  Analysis & Lighting  & Resolution   & Figures \\
        & & & & (mm/px) &  \\
        \hline
        PTVs-01,-02,-03 &  Side  & PTV & Blue LED & 0.8 &   \ref{fig:comp8_radial},\ref{fig:comp8_vertical},\ref{fig:comp7-9_pow} \\
        PTVa-01,-02,-03 &  Above & PTV & Blue LED  & 0.6 &   \ref{fig:comp7-9_pow} \\
        CC-01 &  Above &  CC & White LED & 0.6 &   \ref{fig:cross_corr},\ref{fig:comp4_vel}\\
         PTVs-compare & Side & PTV & Blue LED  & 1.0 &   \ref{fig:comp4_vel} \\
        \hline
    \end{tabular}
    { \footnotesize \begin{singlespace}
\par Notes: PTV stands for particle tracking velocimetry, CC stands for cross correlation.  PTVs-compare is a video used to compare accelerometer data with surface particle displacements. 
Video PTVs-compare and accelerometer placement SR8-compare are of the same impact. 
Videos PTVa-01 and PTVs-01 are of the same impact,  similarly for PTVa-02/PTVs-02 and PTVa-03/PTVs-03.
\end{singlespace} }
\label{tab:exp_videos_list}
\end{table*}

\subsection{Accelerometers and accelerometer placement}

The accelerometers are 5V-ready analog break-out boards by Adafruit\footnote{ \url{https://www.adafruit.com/product/1018}} 
which house a $\pm16$g triple-axis accelerometer (ADXL326) Analog devices integrated circuit. 
The integrated circuit is powered with a 3.3\! V on-board voltage regulator. 
The dimensions of the accelerometer printed circuit boards (PCBs) are 19 mm $\times$ 19 mm $\times$ 3 mm. 
We minimized sensitivity to vibrations in the room by connecting to the PCB with flexible multi-core high gauge wire. 
As done previously \citep{Quillen_2022}, we removed three on-board filter capacitors from the PCB to increase the output bandwidth upper limit from 50 Hz to 1600 Hz on $x$- and $y$-axes and to 550 Hz on the $z$-axis. We only use the $x$ and $y$ axis accelerometer signals because they have a higher bandwidth upper limit. The bandpass upper limits are frequencies at which the signal amplitude is reduced by 3 dB (the amplitude drops by a factor of 0.5) and approximately equal to the cutoff frequency of a low pass filter. The 1600 Hz bandwidth upper limit corresponds to a half period of 0.3 ms which is shorter than the width of the acceleration pulses seen in our experiments.  
The output signals of the $x$ and $y$-axis outputs of the accelerometers were recorded with
 8-channel digital oscilloscopes (Picoscope model 4824A) with a sampling rate of 100 kHz.

Coordinate directions for the experiments differ from those used to describe the accelerometer.   We show cylindrical radius, $\bm{\hat R}$ from
the site of impact and $\bm{\hat z}$, giving height above the surface in Figure \ref{fig:exp_setup}a.  For embedded accelerometers, 
the distance in spherical coordinates from the site of impact $r = \sqrt{R^2 + z^2}$.

Accelerometer locations for the different experiments are summarized in Table \ref{tab:exp_accel_list}. 
Locations are measured from the  
accelerometer integrated circuit which is located at the center of the PCB. 
Accelerometers listed at a depth of zero have half of the integrated circuit embedded into the sand. 

A 67 cm straight metal bar is used to level the surface after raking the sand. The accelerometers are then embedded into the substrate.
We orient the accelerometers so that their $+x$ axes points away from the impact site and their $+y$ axes point vertically up. 
Because it has a lower bandwidth than the accelerometer's other two axes, we don't use the accelerometer internal z-axis channel that is perpendicular to the circuit board plane. 
To ensure that the accelerometers are correctly spaced, are at the desired depth, and are correctly oriented, we individually placed each accelerometer in the sand. 
Long and short tweezers were used to embed the accelerometers. 
The tweezer prongs are marked at centimeter intervals along their length so we can set accelerometer depth. 
The DC voltage levels of each accelerometer were monitored in both axes during placement to monitor their orientation. 
We compared the DC voltage levels of the accelerometer signals prior to impact to the calibration values and find that the accelerometers, once embedded, are typically within $10^\circ$ of the desired orientation.
Calibration values were determined by measuring the voltage along an axis at $\pm$1g and taking the difference.

For one experiment, labeled `PTVs-compare' in Table \ref{tab:exp_videos_list}, we simultaneously recorded the impact with a high speed video and two accelerometers. 
Both accelerometers are located at $R=8$ cm but at depths 0 and 5 cm. 
The accelerometer data is labeled as `SR8-compare' in Table \ref{tab:exp_accel_list}. 
These data are used to directly check the grain motions in time measured from PTV analysis to those measured with accelerometers.
The SZ0-a,-b,-c experiments have 4 accelerometers at different radii but with all accelerometer's integrated circuits on the surface at a depth of 0.
The SZ0-a,b,c experiments are used to corroborate peak velocities measured from the accelerometers to those measured using high speed video in the PTV experiments. 
The SR8-a,-b,-c experiments have three accelerometers at $R=8$ cm and at depths of 0, 2 and 4 cm. 
The SR12-a,-b,-c  experiments similarly have three accelerometers at $R=12$ cm and at depths of 0, 2 and 4 cm. 
The SR8-a,b,c and SR12-a,b,c experiments are used to compare motions near the surface as a function of depth. 
The SR5-ray, SR8-ray, SR10-ray, and SR12-ray experiments each have 5 accelerometers at different depths. 
These experiments are used to study how the velocity vector  direction, or ray angle, vary as a function of position and time and how the surface flow direction is related to the direction of motion below the surface. 
In Table \ref{tab:exp_accel_list} for each experiment we also list the Figures that use measurements from the accelerometer data.

\begin{table*}[htbp] 
\if \ispreprint1
\else\tiny
\fi
\caption{Accelerometer placement coordinates $(r,R, \phi, z)$} 
\label{tab:template}
\begin{center}
\begin{tabular}{llllllllll}
\hline
Experiment  & A          & B           & C              & D             & E  & Figures \\
 \hline 
SZ0-a,b,c & (8.0,8,0,0) & (10.0,10,0,0) & (12.0,12,0,0)  & (14.0,14,0,0)  & & \ref{fig:comp7-9_pow}\\
SR8-compare       & (8.0,8,0,0) & (9.4,8,45,-5)     & & & & \ref{fig:comp4_vel}\\
SR8-a,b,c & (8.0,8,-45,0) & (8.2,8,0,-2)     & (9.0,8,45,-4) &  &  & \ref{fig:skin_delay},\ref{fig:skin_vel}\\
SR12-a,b,c & (12.0,12,-45,0)   & (12.2,12,0,-2)   & (12.6,12,45,-4) &  & & \ref{fig:skin_vel} \\
SR5-ray \!\!  & (5.5,5.5,-90,0) & (5.9,5.5,-45, -2) & (6.8,5.5,0,-4) & (8.1,5.5,45,-6) & (9.7,5.5,90,-8) &\ref{fig:ray} \\
SR8-ray \!\!  & (8.0,8,-90,0)   & (8.2,8,-45,-2)   & (8.9,8,0,-4)   & (10,8,45,-6)   & (11.3,8, 90,-8)  & \ref{fig:ray} \\
SR10-ray \!\! & (10.0,10,-90,0)  & (10.2,10,-45,-2)  & (10.8,10,0,-4)  & (11.7,10,45,-6)  & (12.8,10,90,-8)  &\ref{fig:ray} \\
SR12-ray \!\! & (12.0,12,-90, 0)  & (12.2,12,-45,-2)  & (12.6,12,0,-4)  & (13.4,12,45,-6)  & (14.4,12,90,-8) & \ref{fig:ray} \\
\hline 
\end{tabular}
\end{center}
\par {\footnotesize
\begin{singlespace}
Notes: Each row gives a set or template of accelerometer coordinates.
These are radius $r = \sqrt{R^2 + z^2}$ and  cylindrical coordinates $R,\phi,z$, for each accelerometer with $r,R, z$ in cm and $\phi$ in degrees.
The impact point is at the origin and on the surface.  
The column heads refer to the oscilloscope channels. 
The first column shows names of each experiment. 
SR is for experiments into sand with constant radial position.
The naming of each run is defined by SR + radial distance to the site of impact. 
For example, SR8 stands for the experiments into sand taken at a constant radial distance of 8cm away from the impact site. 
SZ stands for an experiment into sand with all accelerometers at constant depth and is followed by the depth in cm.
The angles ($\phi$) are crudely estimated and should not affect the measurements because of the cylindrical symmetry of our experiment.
SZ0-a,-b,-c refers to three separate experiments.
The SR8-compare experiment recorded the same impact seen in the PTVs-compare video. 
\end{singlespace} 
\label{tab:exp_accel_list}}
\end{table*}

\subsection{PTV particle tracking}

To track the displacement of the fluorescent grains with respect to time we use the Python package \texttt{Trackpy} \citep{trackpy}. 
\texttt{Trackpy} is a software package for finding blob-like features in video, tracking them through time, linking, and analyzing the trajectories. It implements and extends in Python the widely-used Crocker-Grier algorithm for finding single-particle trajectories \citep{crocker96}.
Because the millet particles are relatively large ($\sim$ 3 mm in diameter), prior to tracking them, we use Gaussian blur and erosion spatial filters on each image frame. This improved the accuracy of the centroid algorithm.   Blurring was not needed for tracking individual fluorescent sand particles. 
To calculate the velocity as a function of time we filter the displacements with a Savitsky-Golay filter prior to computing the time derivative.

The videos showing the side view of the impact were used to measure radial and vertical displacements.  The videos taken from above the impact were used to measure radial displacements only. The radial measurements from both viewpoints were consistent. We indicate the radial direction in cylindrical coordinates by 'R' and the vertical direction by 'z'.

\subsection{Measuring surface displacements with cross-correlation}
\label{sec:cross}

The cross-correlation technique we use is similar to that described by \citet{Honda_2021} in their section 2.3.
We start with a video frame prior to impact, that is converted into a gray scale image, $I_0$.
The horizontal and vertical directions were the radial $\bf \hat R$ and azimuthal angle \bm{$\hat \phi$} axes.
We extract a $64 \times 64$ pixel sub-image (equal to $3.58 mm \times 3.58 mm$) from $I_0$,  which we use as a template $T_0$. 
We subtract the mean pixel value from the template image.
$T_0[i,j]$ gives the gray-scale intensity of the template pixel indexed by $i,j$. 
We choose a subsequent video frame, which we refer to as $I_1$, and from it subtract its mean value. 

We compute a cross correlation image 
\begin{equation}
R_{CC}(\Delta i, \Delta j) = \sum_{i,j} T_0[i,j]
I_1[i+\Delta i, j+ \Delta j].
\end{equation}
by taking our template $T_0$ and shifting it across the entire image along one axis, the radial axis.
The location of the peak value in $R_{CC}$ gives the relative shift between template $T_0$ and video frame $I_1$.    
The pixel scale (0.056 mm/pixel) is used to convert the horizontal shift $\Delta i$ in pixels to a horizontal displacement in mm.

Template images are extracted at evenly spaced locations within $I_0$, the pre-impact frame.
We computed cross correlation coefficients for a horizontal row of template locations and in a series of video frames that are at evenly spaced times after impact.  The horizontal row extends along the radial direction from impact. 
The distance between template locations is 10 pixels, or $\sim$0.56 mm and the time between the video images was 3 frames which is equivalent to $\sim$0.67 ms.
The result is a measurement of the surface displacement as a function of both radial position and time.  As each measurement is done 
for a $3.58$ mm $\times 3.58$ mm template image, 
the displacement is an average value for particles within this region.  
Displacements measured via cross correlation will be discussed further in section \ref{sec:CC}. The cross correlation, PTV analysis, and accelerometer methods are compared in section \ref{sec:compare}. 

\section{Results on Surface Displacements and Velocities}
\label{sec:PTV}

\subsection{Particle Tracking Velocimetry (PTV)}

\begin{figure*}[htbp]
\centering
\includegraphics[scale=0.7]{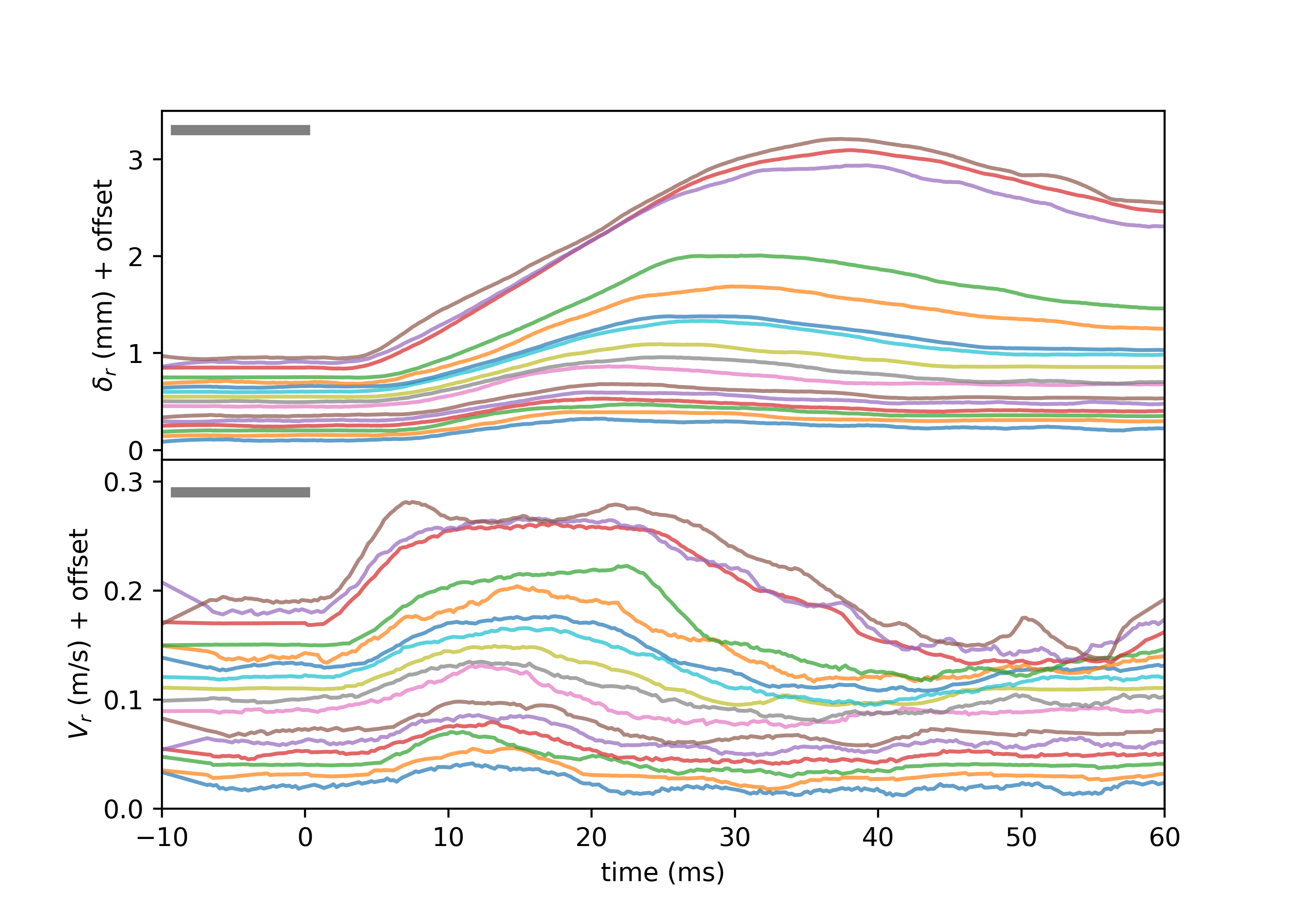}
\caption{
Radial particle displacement (top panel) and velocity (bottom panel) as a function of time for different particles tracked in the PTVs-01 video. Each line shows the trajectory of a single fluorescent millet or sand particle. 
The displacements and velocities are filtered with a Savitsky-Golay filter with window duration, $\sim$10 ms, indicated by the gray bar in the top left of each plot.
Vertical offsets have been added to each line to separate them. 
Curves are in order of distance from the impact, with the top curves nearest the site of impact.  Grains initially range from $R = $ 7 to 14 cm from the site of impact.  
}
\label{fig:comp8_radial} 
\end{figure*}

In Figure \ref{fig:comp8_radial} we show radial displacements of fluorescent millet and sand surface grains as a function of time from the PTVs-01 video. The radial displacements for videos PTVs-02,03 are similar.
The top panel shows radius $r$ plus an offset and the bottom panel shows velocity $v_r$ plus an offset.  
Because particles are on the surface cylindrical coordinate $R \approx r$, where $r$ is distance to the impact point of first contact.
Each particle is shown with a different color line. 
The vertical offsets are used to separate the particle trajectories so that they can be more clearly compared.
Particles are plotted in order of distance from the point of impact with those nearest the point of impact on the top.  
The horizontal axis shows time from the moment of impact.  The time of impact was identified by finding the video frame where the projectile first touches the substrate surface. 
The displacements and velocities were smoothed by a Savitsky-Golay filter. The filter window is displayed with a gray bar on both panels. 
The filter width, $\sim10$ ms, was adjusted so that it did not significantly change the magnitude or timing of the pulses.
Particle trajectories are only plotted before they are obscured by or affected by ejecta. 

\begin{figure*}[htbp]
\centering
\includegraphics[scale=0.7]{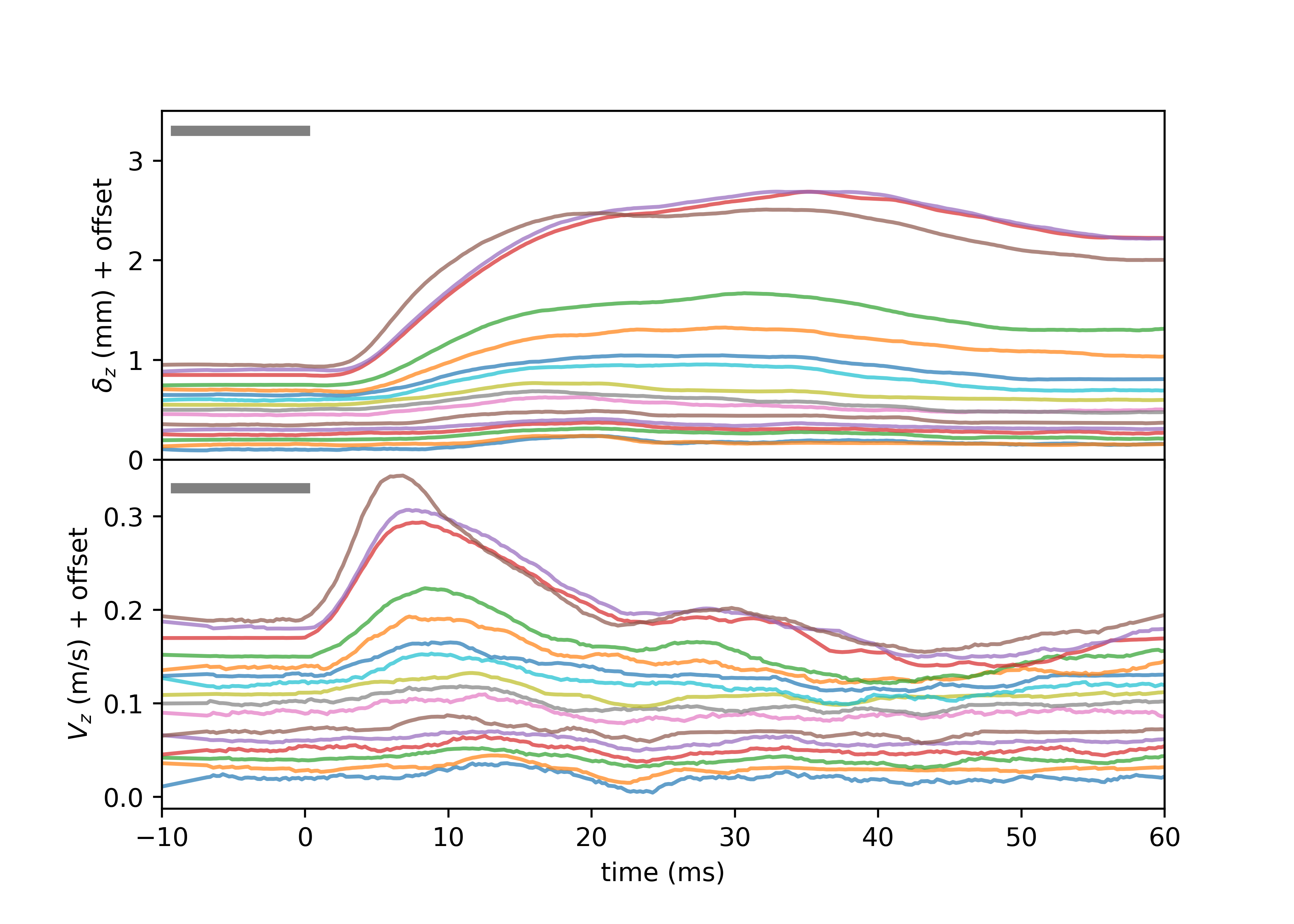}
\caption{ 
Similar to Figure \ref{fig:comp8_radial} except showing vertical displacement and velocity. }
\label{fig:comp8_vertical} 
\end{figure*}

\begin{figure}[htbp]
\centering
\includegraphics[scale = 0.6]{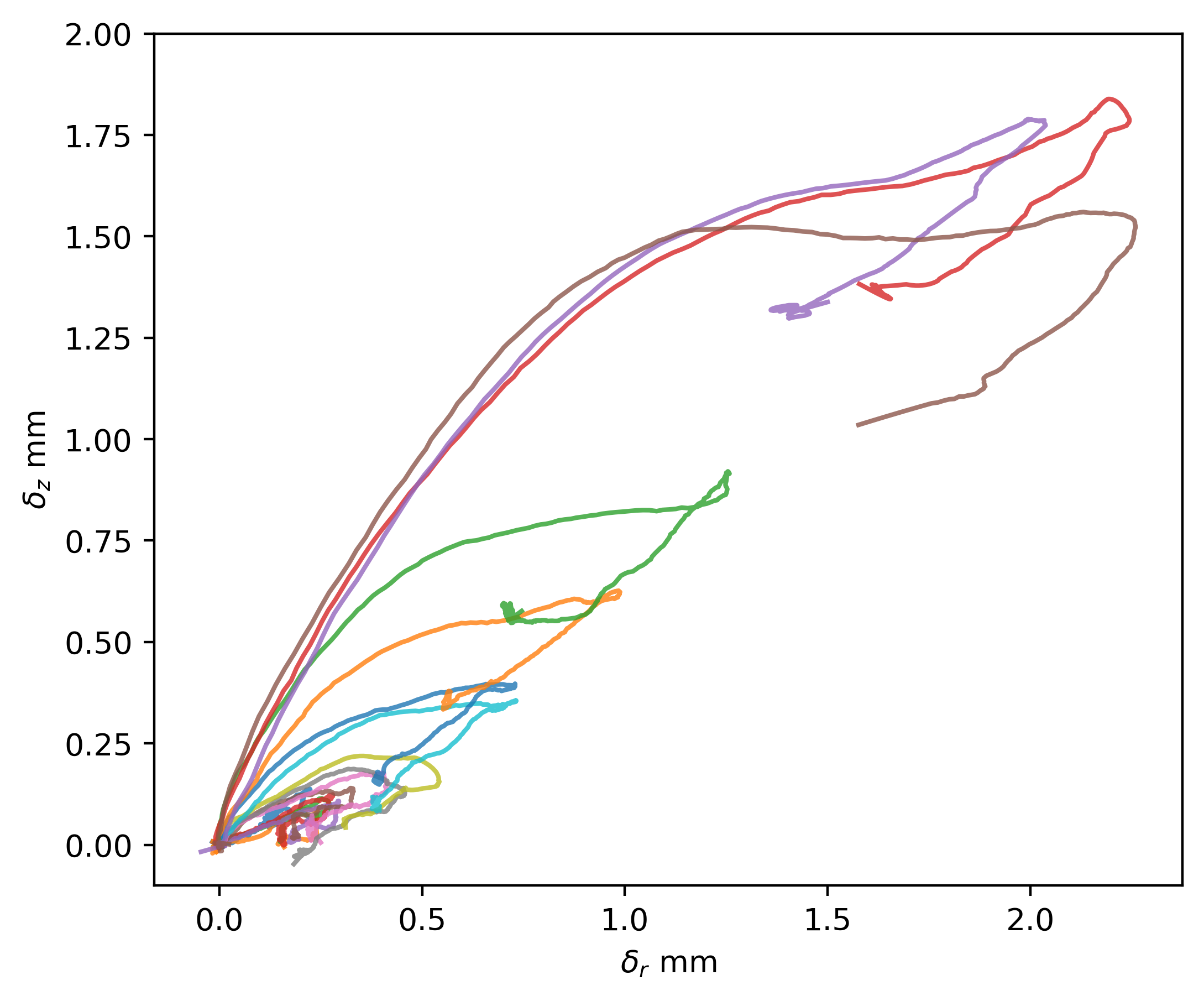}
\caption{Vertical versus radial displacement for the particles depicted in Figures \ref{fig:comp8_radial} and \ref{fig:comp8_vertical}. Each line is plotted as the coordinate pair ($\delta_r(t)$, $\delta_z(t)$) for $t<$ 60 ms. The displacements are filtered with a Savitsky-Golay filter with window duration of $\sim$10 ms. The tracked particle initial positions range from $R=7$ to 14 cm from the site of impact.
}
\label{fig:zvx}
\end{figure}

The top panel of Figure \ref{fig:comp8_radial} 
shows that the particles' final radii are larger than their initial values.  This means that surface particles are permanently displaced after the impact, outward and away from the point of impact.  
Figure \ref{fig:comp8_vertical} is similar to Figure \ref{fig:comp8_radial} except the vertical displacements and velocities are plotted. 
The top panel shows that particles are permanently lifted by the impact. 
We refer to the difference between final and initial positions as final displacements and denote their radial and vertical components as $\delta_{r,f}$ and $\delta_{z,f}$, respectively.  
Figure \ref{fig:zvx} plots $\delta_z$ versus $\delta_x$ for the same particles in figures \ref{fig:comp8_radial} and \ref{fig:comp8_vertical}. 

Figures \ref{fig:comp8_radial}, \ref{fig:comp8_vertical}
and \ref{fig:zvx} show that peak radial and peak vertical displacements exceed the final displacements. The initial radial distance of the particles ranges from 7 to 14 cm from the site of impact.
The surface exhibits some elastic behavior or relaxation as it springs back, both radially and vertically, after reaching a maximum displacement. 
Because particles do not return to their initial locations, there is plastic or ductile deformation or flow. 
The peak radial displacement is about $\delta_{r,pk} \approx 2$ mm for particles at $r=7$ cm which is about 1.2 $R_c$. The peak vertical displacement is about 
$\delta_{z,pk} \sim 1.5$ mm at the same radius.

The width of the radial velocity component signal in Figure \ref{fig:comp8_radial} is about 30 ms for the particles closest to the impact site. 
The width of the vertical velocity component signal in Figure \ref{fig:comp8_vertical} is about 15 ms.
These durations are 5 to 10 times longer than the 2 ms duration (measured with a FWHM) of subsurface velocity seismic pulses observed in similar experiments at 5 cm below the surface \citep{Quillen_2022}. 
Using crater scaling arguments,
\citet{Housen_1983} estimated that crater excavation takes place on a time-scale 
\begin{equation}
\tau_{ex}=\sqrt{\frac{R_c}{g}}.
\label{eqn:time}
\end{equation}
For our experiments and using values listed in Table \ref{tab:exp_constants} we estimate $\tau_{ex}= 75$ ms.
The long duration of the velocity pulses is closer
to the expected crater excavation time-scale than
the duration of the subsurface pulses.

\subsection{Peak and final surface displacements}
\label{sec:peak}

\begin{figure}[htbp]
\centering
\if \ispreprint1
\includegraphics[scale = 0.52]{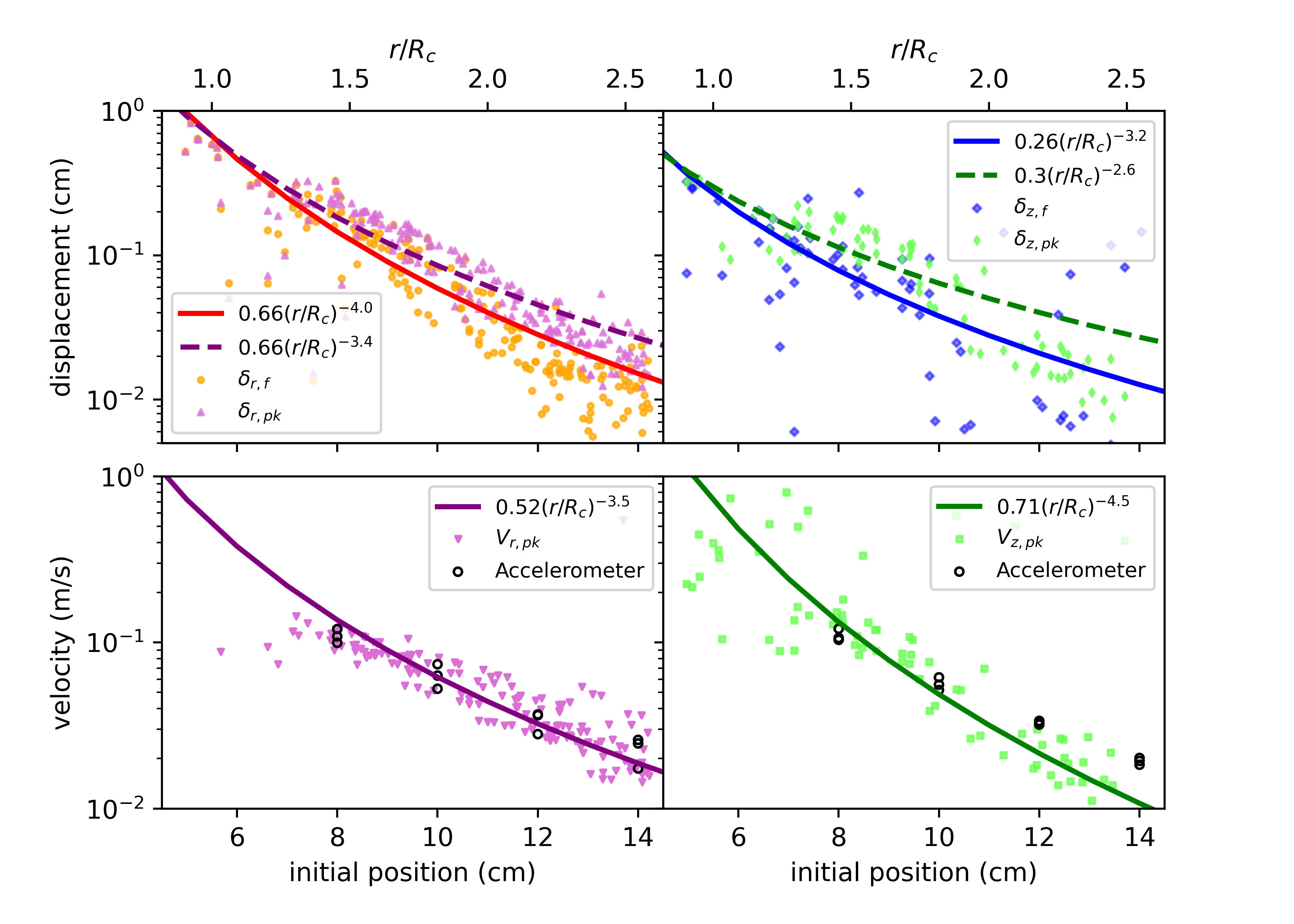}
\else
\includegraphics[width=5in]{power_law2.png}
\fi
\caption{
Radial and vertical displacements as a function of distance from the point of impact along with power law fits.  
This figure uses grain displacements measured and calculated velocities from videos PTVa-01,-02,-03 and PTVs-01,-02,-03. 
Velocities calculated from surface accelerometer measurements from experiments SZ0-a,-b and -c are shown as black circles.
Top left:  Final radial displacements are shown with orange dots. Peak displacements are shown with purple triangles.
The curves show power law fits to each data set. 
Top right:  Final vertical displacements are shown with blue diamonds. Peak vertical displacements are shown with thin green diamonds.
Lower left: Peak radial velocities are shown as purple triangles.
Lower right:  Peak vertical velocities are shown as green squares. 
Note that the large amount of scatter in the vertical data (upper right) is due to ambiguity in the camera viewing angle.
Exponential fit parameters for these data can be found in Table \ref{tab:exp_fits}.
}
\label{fig:comp7-9_pow} 
\end{figure}

From the tracked fluorescent particles from six  
videos PTVa-01,02,03 and PTVs-01,02,03, we measured peak radial $\delta_{r,pk}$ and vertical $\delta_{z,pk}$ displacements and final radial $\delta_r$ and vertical $\delta_z$ displacements as a function of initial particle radius from the site of impact.  
Videos PTVa-01,02,03 were used to  measure radial displacements and PTVs-01,02,03 videos were used to measure both radial and vertical displacements.   We also computed and measured peak radial and vertical velocity components
$v_{r,pk}$ and $v_{z,pk}$ from the same tracked particles. 
In Figure \ref{fig:comp7-9_pow} we plot the peak and final radial displacements in the top left panel, peak and final vertical displacements in the top right panel. Peak radial velocities are plotted in the bottom left panel, and peak vertical velocities in the bottom right panel.  
The vertical axes are on a log scale.   The horizontal axes on the bottom show radial distance from the site of impact in cm and on the top they show distance in units of the crater radius $R_c$. 

Independently, we estimate the velocity of surface motions by integrating the accelerometer signals from 4 near surface accelerometers located at different radii in experiments SZ0-a, -b, and -c.  Peak radial and vertical velocities from the accelerometer signals are plotted with black round dots in the bottom two panels of  Figures \ref{fig:comp7-9_pow}.
The peak velocities computed from the accelerometer signals agree with those computed from tracking surface particles. 
In Figure \ref{fig:comp7-9_pow} we notice that measurements from six videos of three different impact experiments are consistent with the accelerometer measurements.    

The decay of subsurface seismic pulse properties as a function of distance from the site of impact have been fit with a power-law  \citep{yasui15,Matsue_2020,Quillen_2022}.  
In contrast \citet{Honda_2021} fit an exponential function to crater rim height as a function of radial distance for Hayabusa2's artificial impact. In response, we fit two different functions to the peak and final displacements and peak velocities.   

We fit a power law in the form 
\begin{equation}
[X] = A \left( \frac{r}{R_c} \right)^\beta, \label{eqn:powerlaw}
\end{equation}
where $[X]$ is the quantity $\delta_{r,f}$, $\delta_{r,pk}$,  $\delta_{z,f}$, $\delta_{z,pk}$, $v_{r,pk}$
or $v_{z,pk}$.
Fitted parameters are $A$, the variable value at the crater radius $R_c$ and power law index $\beta$.
The power law fits are shown as solid lines in Figure \ref{fig:comp7-9_pow} and the fitted parameters are listed in Table \ref{tab:power_fits}.

We also fit an exponential function in the form
\begin{equation}
    [X] = K \exp \left(-\frac{(r - R_c)}{R_c \lambda}\right).
    \label{eqn:exponential}
\end{equation}
The fitted parameters are $K$, the variable value at the crater radius $R_c$, and a decay length scale $\lambda$ that is in units of $R_c$.
Exponential fits 
and their coefficients are listed in Table \ref{tab:exp_fits}.

The standard deviation for errors in radial position are estimated from the uncertainty of the position of the crater center which is about 0.3 cm. For the error in the particle displacement measurements,  for standard deviation we use a grain radius, which was about $\sigma_\delta \approx 0.02$ cm.  We estimate the error in the velocity components, $\sigma_V$,  from the scatter about the curve of best fit.  Using these uncertainties, best fit parameters were used to compute the reduced
chi-squared, $\chi_\nu^2$, or $\chi^2/(n-2)$ of the fit, where $n$ is the number of data points and 2 is the number of fitted parameters. 
The reduced chi-squared statistics are also listed in Tables \ref{tab:power_fits} and \ref{tab:exp_fits}.

Exponential fits have slightly lower reduced  chi-squared  values that are not significantly lower than those of the power-law fits.  
We find that radial displacements decay more rapidly than vertical displacements.
The final displacements characterize ductile flow or plasticity.  The peak displacement characterize the size of a rebound, which in part could be elastic.  The power-law indices of the final displacements exceed those of the peak displacements.  This implies that the extent of plastic deformation or flow decays faster than the size of the elastic rebound. 

As expected, the fitted lines for peak and final displacements in Figures \ref{fig:comp7-9_pow}  
intersect near $r=R_c$ where particles are ejected. 
The power-law indices indicate that these quantities decay faster with radial distance than predicted by \citet{Quillen_2022}.  
The peak radial displacement index is -3.4 which is steeper than the -2 index predicted for the peak displacement amplitude in the seismic pulse by \citet{Quillen_2022}.
Similarly the peak radial velocity decays with an index of -3.7 which is steeper than the -2.5 predicted for the velocity amplitude in the seismic pulse by \citet{Quillen_2022}.
They found that pulse propagation is not spherically symmetric and that deeper pulses were stronger. Perhaps the -2.5 exponent estimated for peak velocity in subsurface pulse propagation could be reconciled with the -3.5 exponent by using a pulse propagation and flow model that took into account sensitivity to depth or spherical angle. 

Exponential decay lengths range from 0.37 to 0.58 crater radii.  Trends seen for the power-law exponents are correlated with trends in the exponential scale lengths. Steeper decay gives a more negative index and a shorter exponential decay length. 

\begin{table*}[htbp]
\centering
\caption{Power-law fits}
\begin{tabular}{lcccccccc}
\hline
$A(r/R_c)^\beta$ & $A \pm \sigma_A$ & $\beta \pm\sigma_\beta$ & $\chi_\nu^2$ & $\sigma_{\delta}$  & $\sigma_V$ \\ 
      & (cm)   & -  & - & (cm) & (m/s)  \\
\hline 
$\delta_{r,f}$      & 0.66  $\pm$ 0.05     & -4.04 $\pm$ 0.18       & 2.24     & 0.02    &   -     \\ 
$\delta_{r,pk}$  & 0.66  $\pm$ 0.05     & -3.43 $\pm$ 0.15       & 2.85     & 0.02     &   -    \\ 
$\delta_{z,f}$      & 0.26   $\pm$ 0.02      & -3.25 $\pm$ 0.24       & 1.84     & 0.02      &  -   \\ 
$\delta_{z,pk}$  & 0.30   $\pm$ 0.02      & -2.56 $\pm$ 0.19       & 2.60     & 0.02     &   -    \\ 
$v_{r,pk}$       & 0.52  $\pm$ 0.05      & -3.55 $\pm$ 0.14       & 1.24     &   -    &     0.02 \\ 
$v_{z,pk}$       & 0.71 $\pm$ 0.13     & -4.49 $\pm$ 0.27       & 1.24     &   -    & 0.04   \\ \hline
\end{tabular}
\\
{We fit $f(x) = A(r/R_c)^\beta$ 
to both $r$ and $z$ directional components of final displacements $\delta_{r,f}$, $\delta_{z,f}$ and their peak values $\delta_{r,pk}$, $\delta_{z,pk}$, and peak velocities $v_{r,pk}$, $v_{z,pk}$
with data shown in Figure \ref{fig:comp7-9_pow}.
The resulting fitted functions are plotted as lines in the same Figure.   
The column labeled $\chi_\nu^2$ refers to the reduced chi-squared statistic.  
}
\label{tab:power_fits}
\end{table*}

\begin{table*}[htbp]
\centering
\caption{Exponential fits}
\begin{tabular}{lcccccccc}
\hline
$K \exp \left(-\frac{(r - R_c)}{R_c \lambda}\right)$  & $K \pm \sigma_K$ & $\lambda \pm  \sigma_\lambda$ & $\chi_\nu^2$  & $\sigma_{\delta}$  & $\sigma_V$ \\ 
      & (cm)  & -  & -  & (cm) & (m/s)  \\
\hline
$\delta_{r,f}$     & 0.54 $\pm$ 0.03       & 0.37 $\pm$ 0.02       & 2.17         & 0.02    & -        \\ 
$\delta_{r,pk}$  & 0.53 $\pm$ 0.03      & 0.45 $\pm$ 0.02       & 2.74          & 0.02     &  -   \\ 
$\delta_{z,f}$     & 0.25 $\pm$ 0.01       & 0.42 $\pm$ 0.03       & 1.68         & 0.02       &  -   \\ 
$\delta_{z,pk}$ & 0.28 $\pm$ 0.01       & 0.55 $\pm$ 0.04       & 2.33          & 0.02      &  -    \\ 
$v_{r,pk}$ & 0.26 $\pm$ 0.02       & 0.59 $\pm$ 0.02       & 0.89         &  -   &    0.02   \\ 
$v_{z,pk}$ & 0.37 $\pm$ 0.05       & 0.42 $\pm$ 0.02       & 0.82          &  -    &  0.04    \\ \hline
\end{tabular}
\\
{We fit $f(x) = K \exp \left(-\frac{(r - R_c)}{R_c \lambda}\right)$  to  
both directional components of final displacement $\delta_{r,f}$, $\delta_{z,f}$ and their peak values $\delta_{r,pk}$, $\delta_{z,pk}$, and peak velocities $v_{r,pk}$, $v_{z,pk}$. 
}
\label{tab:exp_fits}
\end{table*}

\subsection{Displacements measured with cross correlation}
\label{sec:CC}

\begin{figure}
\if \ispreprint1
\centering\includegraphics[width=\linewidth]{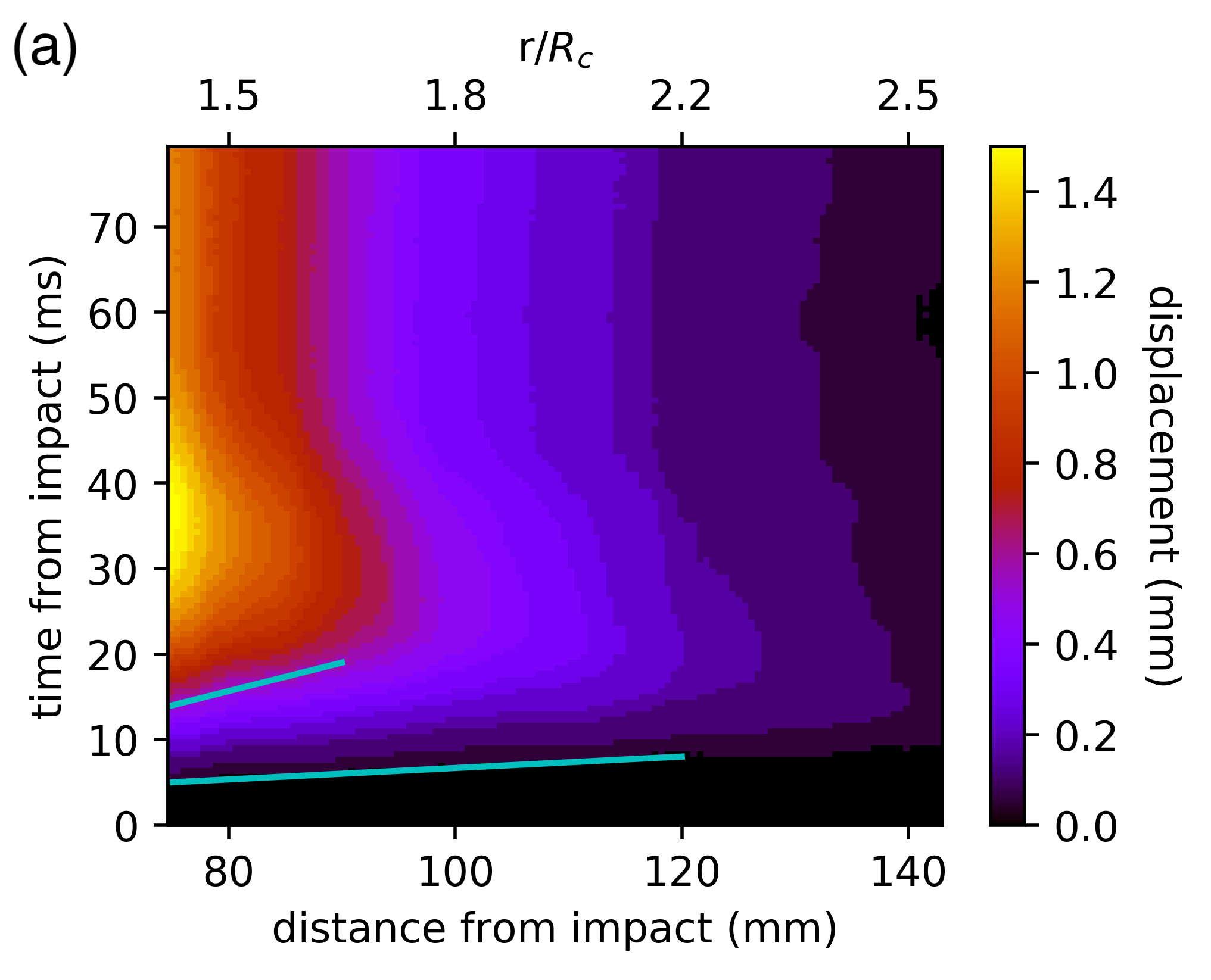} 
\centering\includegraphics[width=\linewidth]{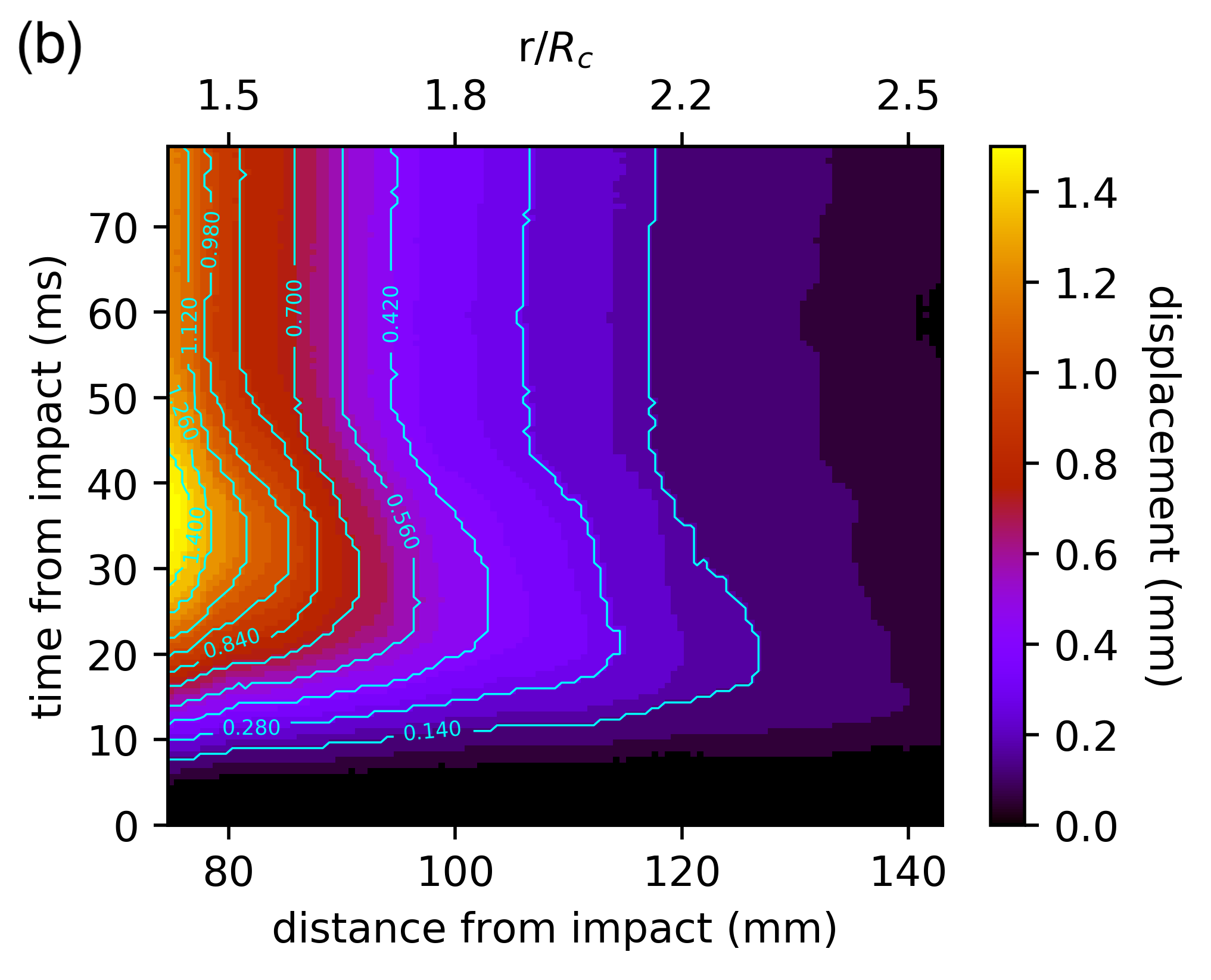} 
\else
\includegraphics[width=3 in]{cross_corr_cc02_slopes2.png}
\includegraphics[width=3 in]{cross_corr_cc02_contours2.png}
\fi
\caption{
Surface displacement as a function of time and distance from time of impact measured from video CC-01 using a cross-correlation method. 
The bottom horizontal axis is the distance from the impact site in millimeters and the top axis is in units of crater radius $R_c$. Vertical axis is time from impact in milliseconds. The color is the amount of displacement in the surface material in units of millimeters.
The surface closest to the impact is displaced the greatest amount. 
Onset of motion is rapid and occurs within 10 ms of impact at radii up to 2.5 crater scale lengths. 
Surface motions show elastic behavior as material  rebounds. 
The surface also shows plastic behavior or flow as final displacements are not zero.
For the cross correlation, distance between template locations is 10 pixels, or $\sim$0.56 mm (in radius), and the time between the video images was 3 frames, which is equivalent to $\sim$0.67 ms.
In (a) we show two lines, with the upper one corresponding to a velocity of 3 m/s and the lower one corresponding to a velocity of 15 m/s. 
Contour lines in (b) shows displacement of surface material.
The displacement profiles have a longer duration closer to the impact site compared to those further away.
This is opposite to the behavior of subsurface pulses which were narrower close to the impact site (as seen in Figure \ref{fig:comp4_vel} and by  \cite{Quillen_2022}).
}
\label{fig:cross_corr}
\end{figure}

Figure \ref{fig:cross_corr} shows surface radial  displacement measured using cross correlation technique in video CC-01, as described in section \ref{sec:cross}.  
The colorbar in Figure \ref{fig:cross_corr} shows the size of the displacement.
The lower horizontal axis shows the distance from the site of impact in mm. The upper horizontal axis is the distance from impact in units of crater radius, $R_c$.
The vertical axis shows time from the first moment of projectile-surface contact.  

Figure \ref{fig:cross_corr} is black on the lower side where there is no displacement at early times. 
Displacement starts to rise at about 5 ms from
the time of impact and rises steeply.  The lower edge shows the travel speed of the initial rise in displacement across the surface.  Within 10 ms of impact, particles at $r=2.5 R_c$ (equivalent to $r=14$ cm) on the surface have started to move. 

The short cyan segment over-plotted on the top panel shows a travel speed of 3 m/s and the longer segment shows a travel speed of 15 m/s. 
The pulse travel speed on the surface is slower than the 50 m/s subsurface pulse travel speed estimated in the same substrate, with similar experiments but using embedded accelerometers \citep{Quillen_2022}.

Displacements do rebound (they decrease at later times) but particle positions do not fully return to their original position which would be near zero.
This confirms that there is plastic deformation or flow beyond the crater radius, as previously inferred from the decay of seismic energy by \citet{Matsue_2020}. 
The image shows a broad pulse, $\sim$20 ms in width, that propagates across the surface.  The pulse duration appears shorter at larger distances from the site of impact.  This too differs from subsurface pulse propagation where pulses (as seen in acceleration or velocity) broadened as they traveled away from the impact site \citep{Quillen_2022}.

\section{Near surface dynamics}
\label{sec:surface}

We find that particle movement on the surface differs from movement below the surface in a number of ways. 
The duration of surface motions are longer than we previously saw from subsurface measurements with accelerometers in similar experiments \citep{Quillen_2022}. 
In the same sand substrate and with accelerometers embedded 5 cm deep, and with a lighter and slower projectile impact, pulse duration in velocity (as a FWHM) were only a few ms \citep{Quillen_2022}, whereas here we find that surface particles just outside the crater radius move for more than 40 ms.   
In section \ref{sec:peak} we found that surface particle peak velocity decays more rapidly with distance from impact site than did the seismic pulse velocity 5 cm below the surface. 
Ray directions at the time of peak acceleration are nearly radial for the subsurface seismic pulses \citep{Quillen_2022}, whereas here we find that surface particles initially move upward at an angle of about $\sim 45^\circ$ from horizontal. 
The onset of motion on the surface travels at about 15 m/s rather than 50 m/s for the subsurface pulse. 
At larger distances from impact, pulse duration on the surface is shorter than nearer the crater and this is opposite to the trend exhibited by subsurface pulses. 

We consider some possible explanations for these differences.   
Firstly, velocity is a derivative of displacement and acceleration is a derivative of velocity.
Pulses seen in displacement can be consistent with those seen in acceleration or velocity but would be wider.  This effect is illustrated in Figure \ref{fig:derivatives}.
Secondly, tracked surface particles might slide or roll across the surface and so they could differ from the bulk of the material next to them.    Alternatively, the motions of substrate material could be sensitive to depth.  This would be supported by recent simulations of impacts that found that plasticity (as measured from particle strains) was confined to a shallow surface layer \citep{Miklavcic_2022}.

\subsection{Comparing particle and accelerometer motions}
\label{sec:compare}

\begin{figure}[htbp]
    \centering
    \if \ispreprint1
    \includegraphics[width = 3.5 in, trim = 30 0 0 0, clip]{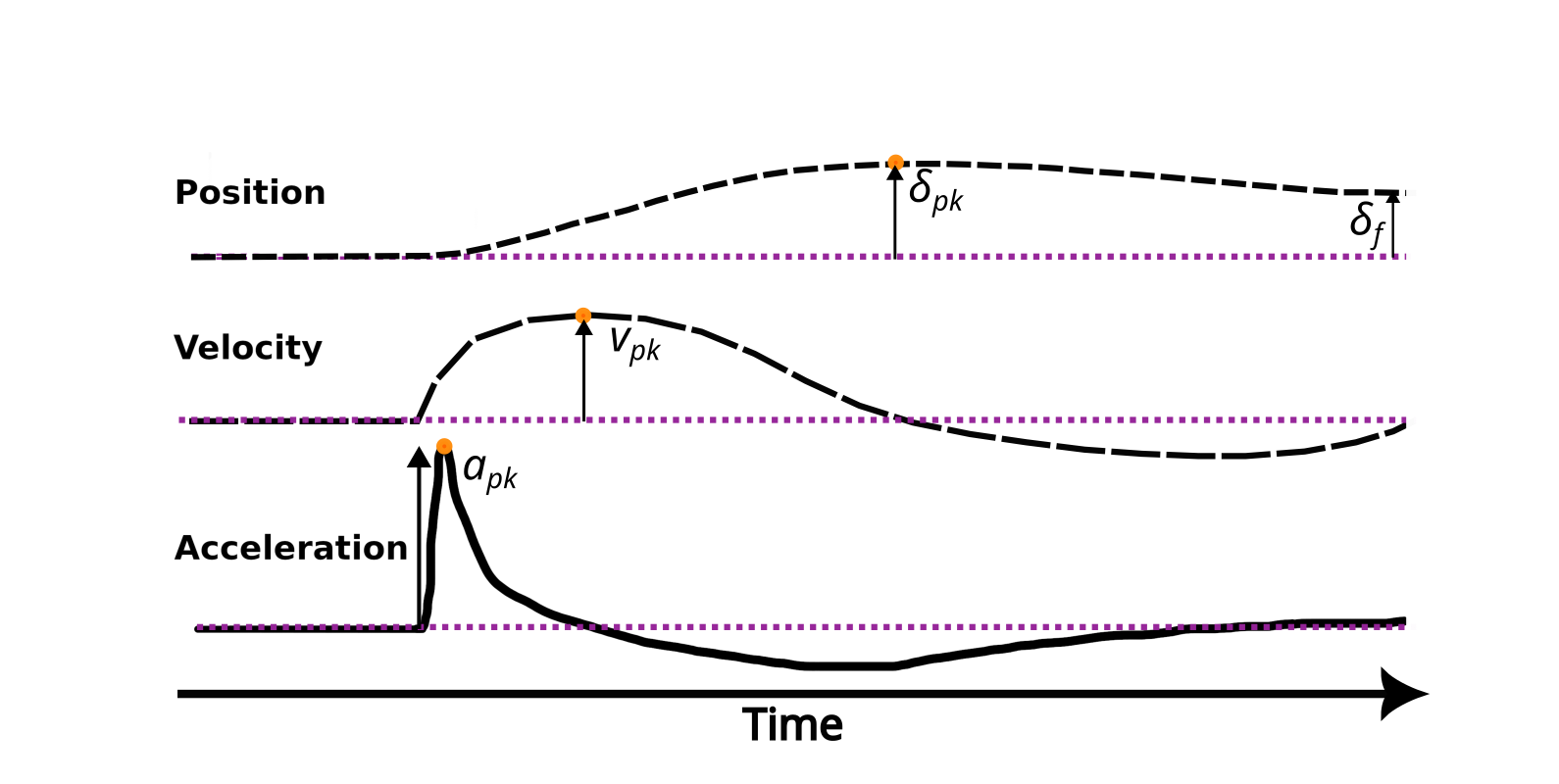}
    \else
    \includegraphics[width = 4.5 in, trim = 30 0 0 0, clip]{derivatives.png}
    \fi
    \caption{Illustration of position, velocity, and acceleration curves resulting from a typical accelerometer measurement. Notice how the pulse broadens and the pulse peak shifts with each integration from the acceleration pulse. 
    }
    \label{fig:derivatives}
\end{figure}

\begin{figure}[htbp]
\centering 
\if \ispreprint1
\includegraphics[scale = 0.45]{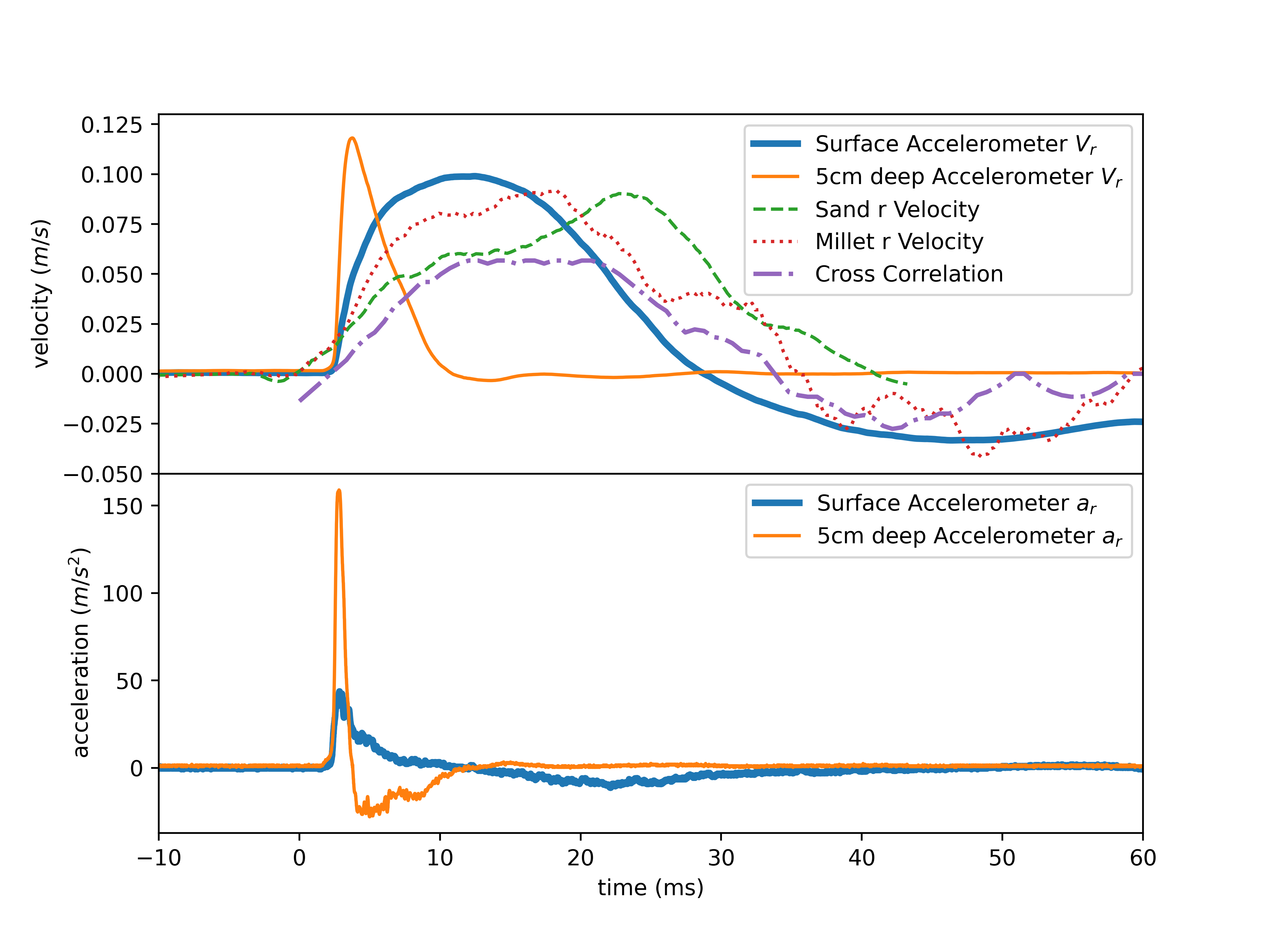}
\else
\includegraphics[width = 4.5 in]{comp4_accel2.png}
\fi
\caption{Top panel: Comparison of velocities computed from accelerometers and videos using PTV and cross correlation methods. The blue thick solid and thin orange lines are radial velocity components computed from accelerometers at $R=8$ cm and at depths of 0 and 5 cm, respectively using the data from the PTVs-compare experiment.  Radial velocity components based on tracking a fluorescent millet grain and a fluorescent sand grain, both at $R=8 $ cm, are plotted with thin dotted red and dashed green lines, respectively. The violet dot-dashed line shows the radial velocity component extracted at $R=8$ cm from the displacement image (see Figure \ref{fig:cross_corr}) that was computed from video CC-01 via cross correlation method.
Surface motions seen in the two surface grains, via cross correlation and in the near-surface accelerometer are similar and consistent.
The deeper accelerometer has a much narrower velocity pulse.  
Bottom panel: The radial component of acceleration is shown as a function of time for the same two accelerometers shown in the top panel.  The deeper accelerometer also has a narrow acceleration pulse. 
}
\label{fig:comp4_vel}
\end{figure}

We compare motions of particles on the surface that are measured using particle tracking techniques to those measured from accelerometers. 
To reduce the accumulated error in our comparison between the accelerations from the accelerometers and tracked particle displacements from the videos, we compare velocities. 
We calculate the velocity by differentiating displacement of the surface particles and by numerically integrating acceleration measured from the accelerometers. 

In Figure \ref{fig:comp4_vel} we compare the radial velocity of a tracked sand and millet particle at $r=8$ cm  and two accelerometers at the same radius from the PTVs-compare and SR8-compare experiment. The accelerometers are at depths of 0 and 5 cm. In contrast, \citet{Quillen_2022} had accelerometers at depths 5 cm or deeper.
This figure also shows velocity at the same radius computed using cross correlation from the CC-01 video. 
We extracted the displacement signal at $r=8$ cm from a vertical line using the displacement array shown in Figure \ref{fig:cross_corr}.

The top panel of  Figure \ref{fig:comp4_vel} shows that the duration of the velocity pulse seen in the near surface accelerometer is long and similar to the durations of the particle velocity pulses measured via particle tracking velocimetry.   
We find that bulk motions measured with the near surface accelerometer are similar to the particle motion on the surface.  
Similarly, particle motions detected via particle tracking velocimetry are consistent with those measured via cross correlation.
Since we are directly comparing velocities, 
differentiation or integration does not account for the difference in pulse widths (as illustrated in Figure \ref{fig:derivatives}).
However,  Figure \ref{fig:comp4_vel} shows that 
velocity pulse duration is shorter in the 5 cm deep accelerometer than in the one on the surface. 
We note that the surface accelerometer effectively averages motion over the top cm of the surface due to the size of the circuit board so we do expect some differences between the accelerometer and surface particle motions.  
At a depth of 5 cm, the radial velocity pulse has a width $\sim$ 5 ms, whereas on the surface the pulse width is $\sim $ 30 ms.  The comparison of the surface particle and accelerometer motions shows that the surface motions are significantly different from sub-surface motions only 5 cm deeper.  The bottom panel of Figure \ref{fig:comp4_vel}, showing radial accelerations for the same two accelerometers, shows that accelerations are similarly sensitive to depth. 
We find that motions near the surface are sensitive to depth.  The differences between impact generated surface motions detected outside the crater and subsurface seismic motions arise because motions are strongly dependent on depth in the top few centimeters. 
 
\subsection{Velocity as a function of depth}

In the previous section we showed that particle motions measured via PTV and cross correlation are consistent with each other and those measured from near surface accelerometers. 
However, we find that velocities measured from accelerometers is sensitive to depth in the top few cm.  
To investigate the sensitivity to depth further we use two sets of three accelerometers placed at depths of 0 cm, 2 cm, and 4 cm.  In the first set of three experiments, denoted SR8-a,-b,-c, the three accelerometers are at radial distances of $R=8$ cm.   The second set of experiments denoted SR12-a,-b,-c is similar except the accelerometers are at $R$=12 cm.  

\begin{figure}[htbp]
    \centering
\if \ispreprint1
    \includegraphics[scale = 0.5]{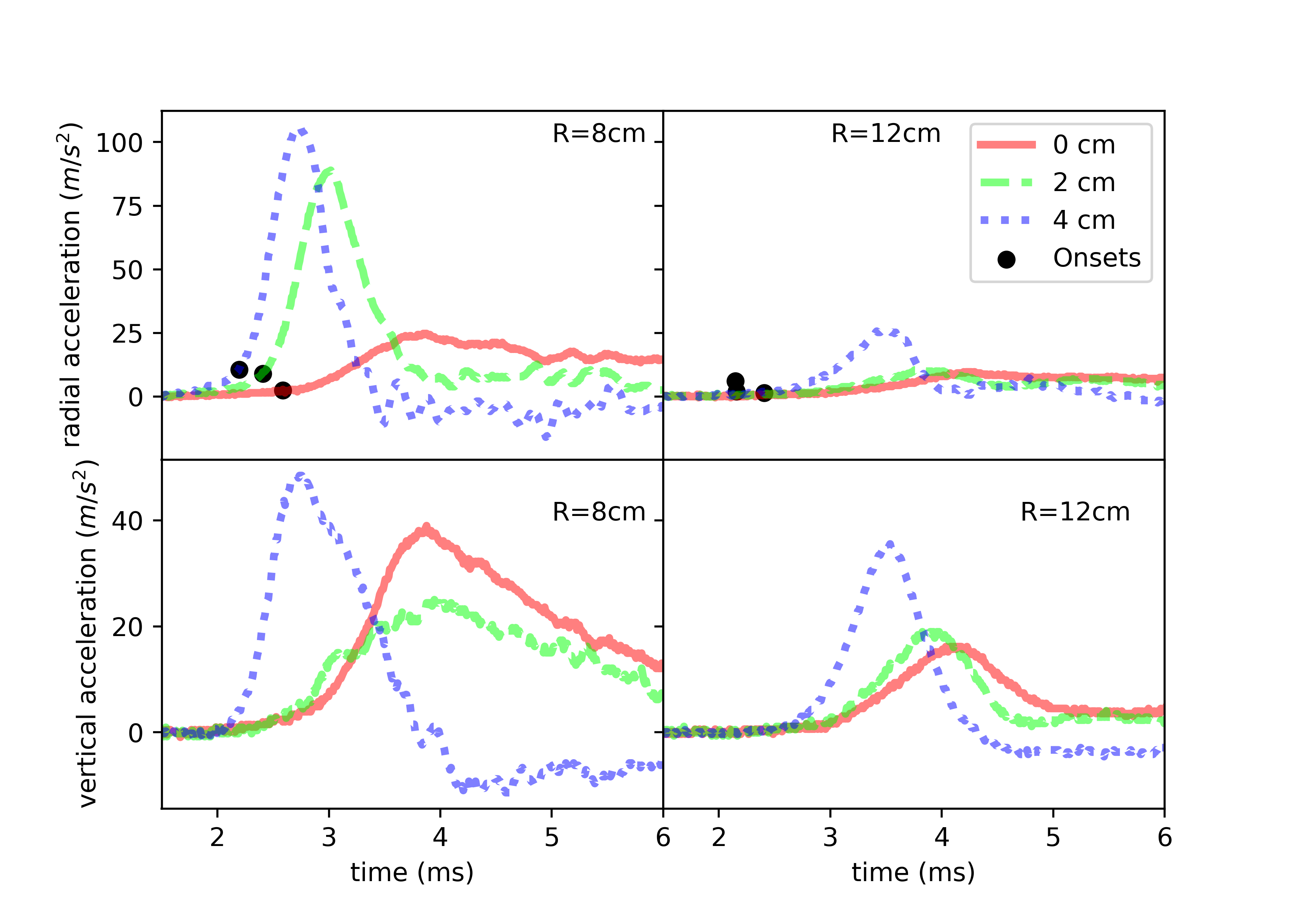}
\else
\includegraphics[width= 4.5in]{skin_delay_tot.png}
\fi
    \caption{Acceleration measured at two different radial distances from the SR8-b  and SR12-b experiments. The top two panels show radial $\hat {\bf R}$ components of acceleration and the bottom two show vertical $\hat {\bf z}$ components. The left two panels each shows acceleration at a radial distance of $R=8$cm from the SR8-b experiment and the right two panels shows acceleration at $R=12$cm from the SR12-b experiment.  The depth of the accelerometer sets the color of the line and the line style.  Pulse shape is strongly dependent on depth. We define the onset of each pulse as the point where the pulse reaches 10\% of the pulse peak value. For the radial components, the onsets of each pulse are plotted using black dots. Pulse onset and peak times are earlier below the surface. 
    }
    \label{fig:skin_delay}
\end{figure}

Figure \ref{fig:skin_delay} shows the radial and vertical (in cylindrical coordinates) components of acceleration for three accelerometers from the experiment SR8-b and SR12-b. We compute pulse onsets by taking the first point where the radial pulse rises above 10\% of the pulse maximum.  Pulse onsets are shown with black dots. 
Figure \ref{fig:skin_delay} shows that with increasing accelerometer depth, the radial pulse not only decreases in duration, but also arrives sooner. The pulse arrives at the $R=8$ cm accelerometer at a depth of 0 cm about 0.5 ms after it arrives at an accelerometer located 4 cm below the surface, despite both accelerometers being at the same radial distance. The pulse arrives at the 2 cm deep accelerometer between the 4 cm and 0 cm arrival times. The pulses arrive a few ms after the first moment of impact and the pulse durations, as seen in acceleration, are only a few ms wide. We also see the pulses attenuate and change shape with radial position. We attribute the relatively low vertical acceleration seen at depth 2 cm in the SR8-b experiment to sensitivity to azimuthal angle, as the three accelerometers are not at the same azimuthal angle. This could arise due to irregularities in the surface level, bed preparation or an off-center impact. 
We note that the bandwidth of the accelerometers is 1500 Hz, which means the onset time differences we detect are at the edge of the accelerometer's capabilities.  

A possible explanation for the sensitivity of pulse arrival time on depth is that pulse travel speed, $v_P$, is sensitive to depth. 
Hertzian contact models predict a power-law dependence of the effective pulse travel speed $v_P$ in a granular medium on ambient or confinement pressure $P_0$ \citep{Duffy_1957,Liu_1992,Johnson_2000,Somfai_2005}
with a scaling of $v_P \propto P_0^{\beta_0} $ with index $\beta_0\approx \frac{1}{6} $. Experiments measure $\beta_0$ in the range $\frac{1}{4}$ to $\frac{1}{6}$ \citep{Tell_2020,Zhai_2020}.  Hydrostatic pressure can act like a confinement pressure, giving $P_0 = \rho g |z|$ and a pulse propagation velocity that decreases near the surface. But because hydrostatic pressure approaches zero near a free surface, the propagation velocity would vanish at the surface.  In this limit it is difficult to predict the propagation of waves in the  medium because it is non-linear (e.g., \cite{Rosas_2018}). 
Simulations of oblique impacts suggest that 
there may be a change in material behavior at a particular depth, referred to as a skin depth \citep{Miklavcic_2022}.

Figure \ref{fig:skin_delay} shows that peak acceleration is strongly dependent on depth with peak value increasing as a function of depth.  Peak radial acceleration on the surface is about 1/4 of that seen only 4 cm deeper. Peak vertical acceleration is also lower nearer the surface.  Peak accelerations estimated from the decay rate of subsurface seismic pulses (e.g, \citealt{yasui15,Matsue_2020,Honda_2021,Quillen_2022}) could overestimate the size of the acceleration on the surface. 


\begin{figure}[htbp]
    \centering
\if \ispreprint1
\includegraphics[scale = 0.5]{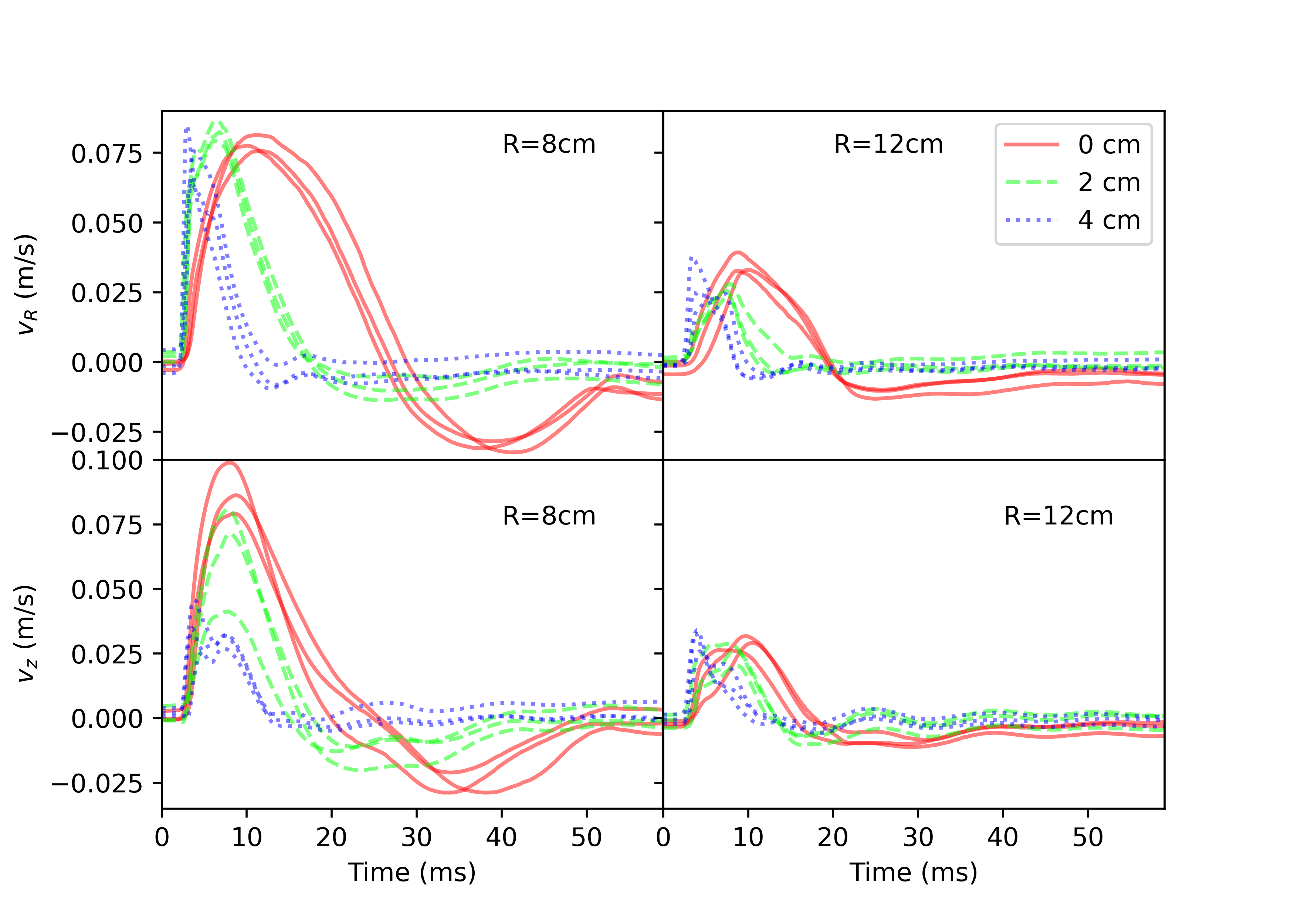}
\else
\includegraphics[width = 4.5in]{skin_vel.png}
\fi
    \caption{Velocities integrated from accelerations measured at two different radial distances from the SR8-a,b,c and SR12-a,b,c experiments. The top two panels show radial $\hat {\bf R}$ components of velocities and the bottom two show vertical $\hat {\bf z}$ components. The left two panels each shows velocities at a radial distance of $R=8$cm and the right two panels shows velocities at $R=12$cm.  The depth of the accelerometer sets the color of the line and the line style. Pulse durations are strongly dependent on depth.}
    \label{fig:skin_vel}
\end{figure}

In Figure \ref{fig:skin_vel} we plot the velocities calculated from all accelerometers in experiments SR8-a,-b and -c and SR12-a,-b and -c as a function of time. We plot radial ($R$) components
in the top two panels and vertical velocity components in the bottom two panels.  
The left two panels show velocities computed using accelerometers at $R=$ 8 cm and those on the right show velocities computed using accelerometers at $R=$ 12 cm.
We see significant broadening on the surface compared to the sub-surface velocity pulses. The pulse widths appear to halve in duration every 2 cm of depth in this 4 cm surface band. This trend is more apparent in the 8 cm radial distance data because the 12 cm radial distance data has more scatter. Since the velocity pulses are of similar amplitude in the radial direction but are much broader on the surface, we conclude that the total momentum flux is higher in the surface layer than in layers just 2 to 4 cm below the surface.  We lack an explanation for the insensitivity of peak velocity and the dependence of pulse duration to depth, however non-linearity of wave propagation in the unconfined surface may be responsible for these phenomena.   

Comparison of Figures \ref{fig:skin_delay} and \ref{fig:skin_vel} shows that velocity pulses are significantly wider than the acceleration pulses.
Because acceleration is the derivative of velocity the acceleration peak is primarily sensitive to the slope in velocity while the velocity first rises. 
The long durations for displacement motions we see on the surface are not simply due to integration because pulse durations seen in both acceleration and velocity signals are strongly dependent upon depth. 

\subsection{Ray angles}
\label{sec:ray}

Using subsurface peak accelerations in similar experiments we found that ray angles were nearly radial from the site of impact \citep{Quillen_2022}. However, the surface particle motions seen here deviate from this expectation as surface particles move outward and upward.
In this section we discuss ray angles for the velocity vectors as a function of position and time and we examine how surface flow is related to the direction of motion below the surface. 

\begin{figure}[htbp]%
 \centering
 \if \ispreprint0
\includegraphics[scale=.55]{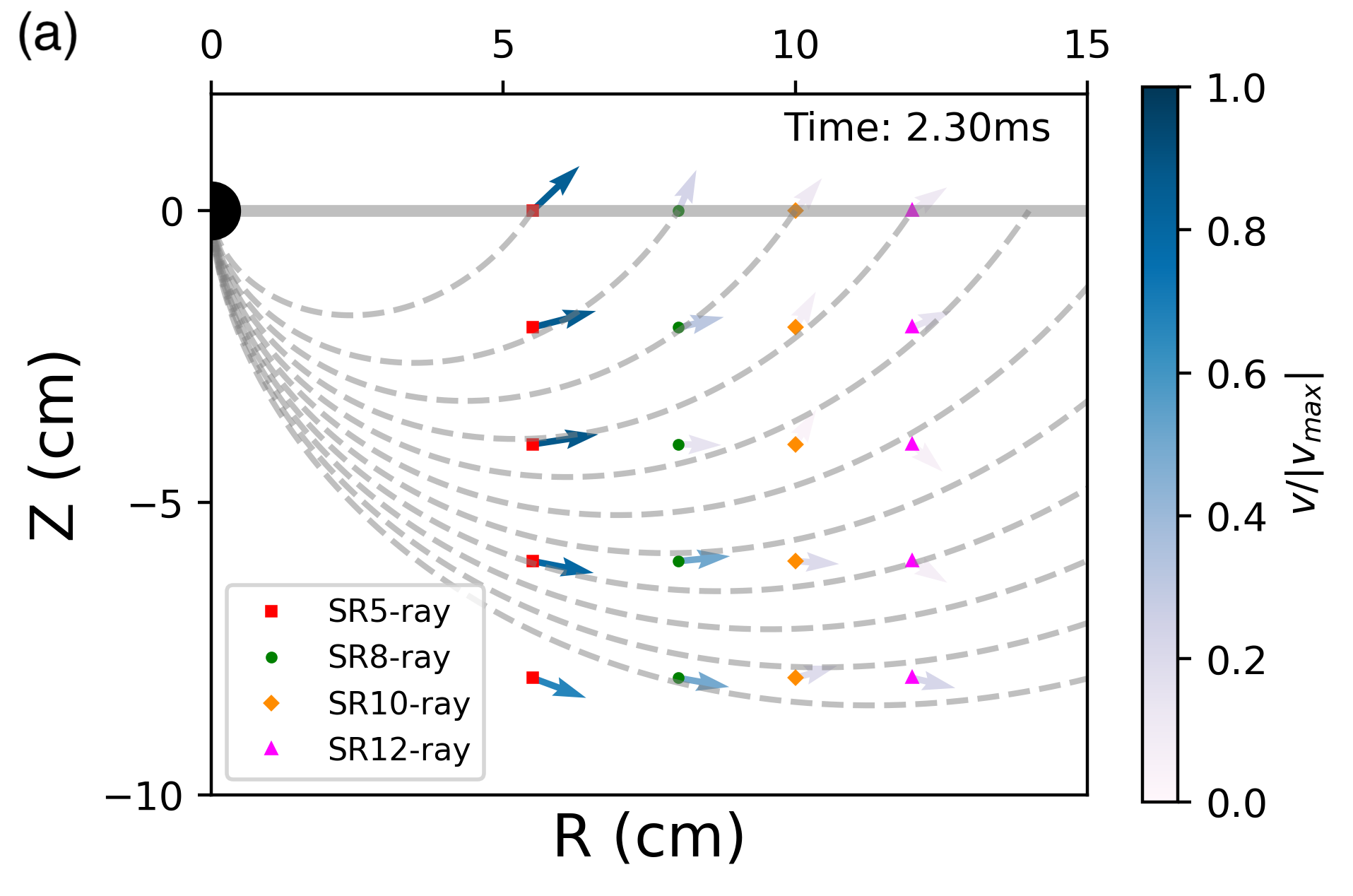}
\includegraphics[scale=.55]{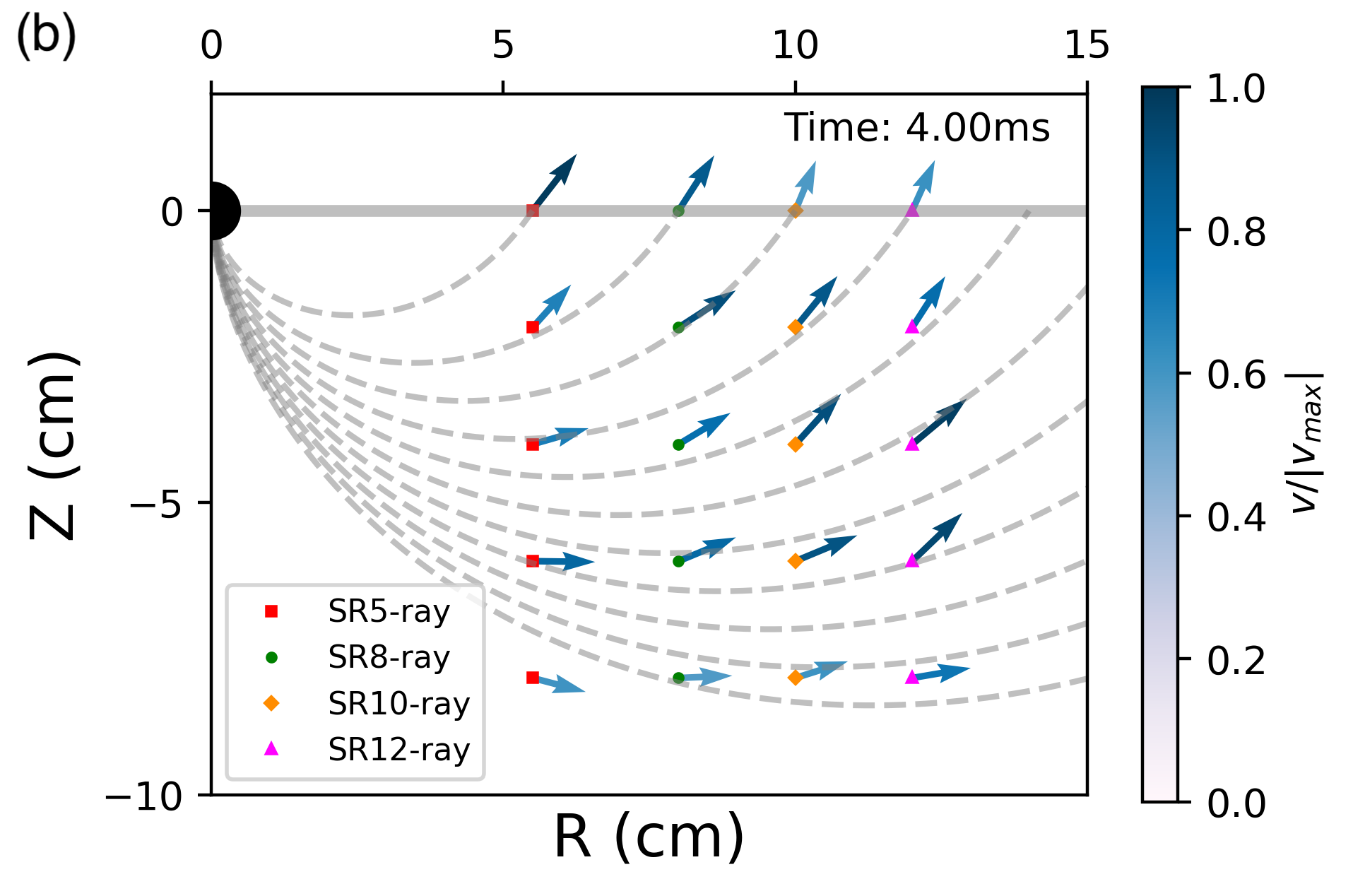}
\includegraphics[scale=.55]{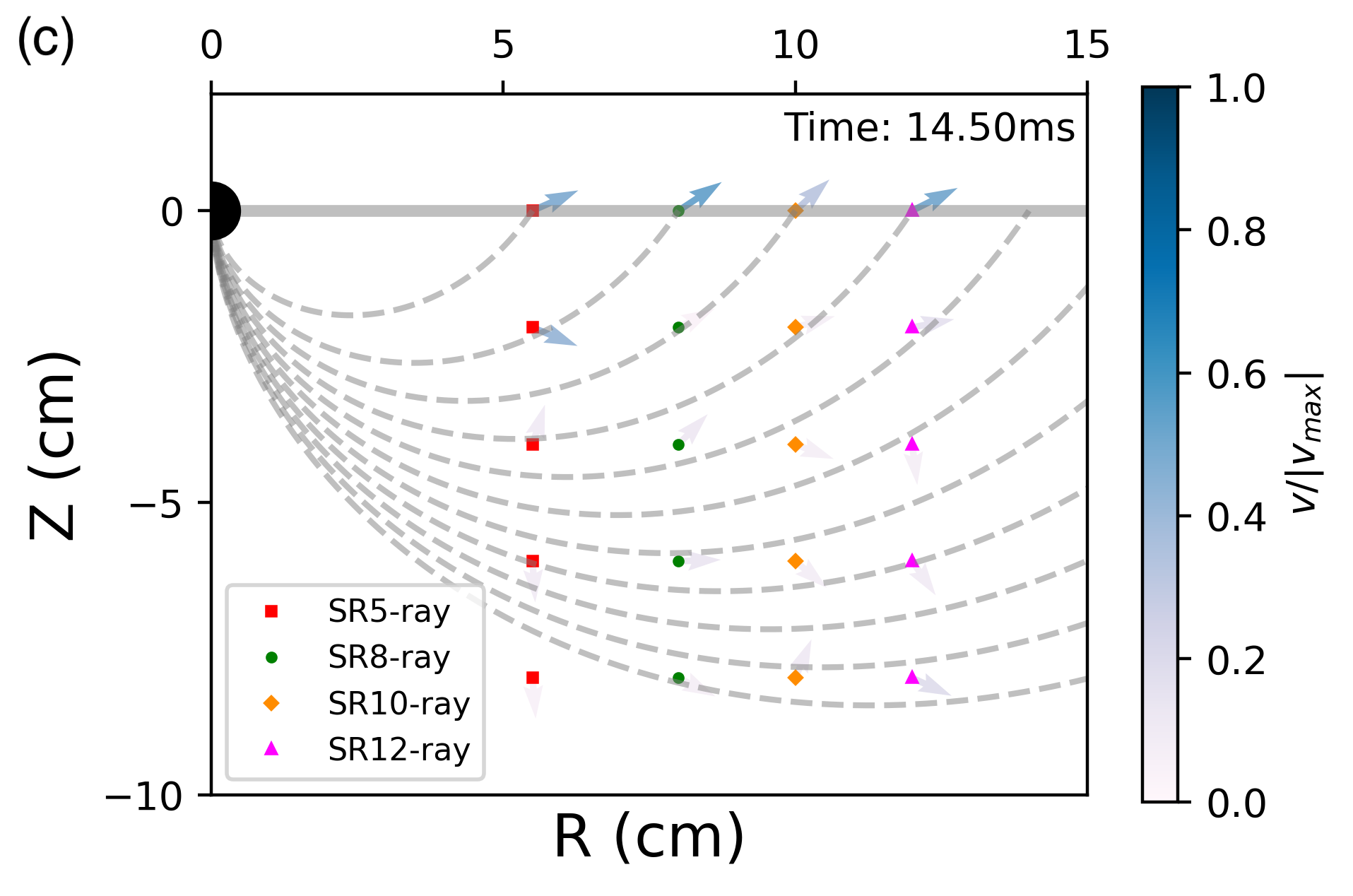}
 \else
    \includegraphics[scale=0.7]{017.png}
    \includegraphics[scale=0.7]{peak.png}
    \includegraphics[scale=0.7]{261.png}
\fi
  \caption{Ray angles for velocity are plotted at three different times. Vectors show velocity direction and are computed using accelerations measured in experiments SR5-ray, SR8-ray, SR10-ray and SR12-ray.  Points show the positions of each accelerometer. The color bar and the length and color of arrows both indicate the magnitude of the velocity vector.  Each color and point type represents a different experiment. The color bar is is given by $|v|/v_{max}$ and is dimensionless. The horizontal gray line $x = 0$ represents the surface of the substrate. The dashed gray lines show streamlines for Maxwell's Z-model with $Z=3.5$ which we gain from the power law decay of the radial velocity in Figure \ref{fig:comp7-9_pow}. Plot axes are in cylindrical coordinates with the site of impact at the origin which is shown with a black solid half circle. The time from the first contact, is shown at the top right corner in each panel.  Motions on the surface have ray angles similar to those below them.  However, motions on the surface continue well after subsurface motion has decayed.   For an animated version of this figure see supplemental video ray\_angles.
  \label{fig:ray} 
  }
\end{figure}

Using signals and coordinates from each accelerometer from experiments SR5-ray, SR8-ray, SR10-ray, SR12-ray,     
we compute velocity by integrating the acceleration. The direction of the velocity is computed using the ratio of the $R$ and $z$ velocity components.
Ray angles are shown with vectors in Figure \ref{fig:ray} at three different times after impact.
In Figure \ref{fig:ray} the length and color axis of the vectors is set by the velocity magnitude. Points and vectors are shown at the position of each accelerometer and each experiment is plotted with a different shape and color point. 
We include a supplemental video, denoted \texttt{ray\_angles}, which shows the ray angles evolving in time. 

When the seismic pulse first arrives at a particular accelerometer, the ray direction is nearly radial, aligned along $\hat {\bf r}$ (and consistent with ray angles measured for peak acceleration by \cite{Quillen_2022}).  
However, soon afterward, the ray rotates so that it points upward.  When the velocity peaks, the flow field resembles that predicted for excavation flows with the Maxwell's Z-model \citep{Maxwell_1977}.
The parameter $Z$ in the Maxwell's Z-model sets the exponent for decay of the radial component of the velocity
\begin{equation}
    v_r(r) = \alpha(t)r^{-Z}.
    \label{eqn:zmodel_z}
\end{equation}
Here $\alpha(t)$ is a time dependent function. 
In spherical coordinates $(r, \vartheta, \phi)$, 
the streamlines for the Maxwell's Z-model obey 
\begin{equation}
    r(\vartheta,R_i) = R_i
    \left(1+\cos{\vartheta}\right)^{\frac{1}{Z-2}},
    \label{eqn:zmodel_r}
\end{equation}
where $R_i$ is the horizontal distance from the impact point to the intersection of a given streamline and the level pre-impact surface (e.g., \cite{Croft_1980,Kurosawa_2019}). The angle $\vartheta$ is defined to start at $0^{\circ}$ from the negative z axis and ranges from $0^{\circ}$ to $90^{\circ}$.


In Figure \ref{fig:ray} dashed gray lines show  streamlines for a Z-model computed with $Z = 3.5$ using Equation \ref{eqn:zmodel_r}.
We chose $Z=3.5$ consistent with the decay rate of the radial component velocity that we measured for peak surface velocities as a function of distance $r$ from the site of impact (shown in Figure \ref{fig:comp7-9_pow} and with power-law parameters listed in Table \ref{tab:power_fits}). 

The similarity between the ray angle directions and the Maxwell's Z-model streamlines suggests that after the initial arrival of the seismic pulse, motion outside the crater excavation region is similar to that within the crater excavation region.  Below the surface, we are seeing a continuation or an extension of the crater excavation flow field. 

Figure \ref{fig:ray} also shows motions detected from accelerometers on the surface. The ray angles 
for these accelerometers match the ray angles below them.   However, 
the subplot (c) in Figure \ref{fig:ray} shows that the surface continues to move well after the subsurface seismic pulse had decayed. The directions of flow on the surface seems connected to the subsurface motion, however, the pulses last much longer on the surface.  One way to account for the duration of motion on the surface is a strongly depth dependent pulse travel speed, with pulse travel speed much slower on the surface than below.  Alternatively the excavation flow within the crater might push material outwards, enhancing the lateral flow near the surface. 
This later interpretation is experimentally supported by the lateral particle motions and high total momentum in horizontal motion observed on the surface.
The similarity of the ray directions on the surface and those below the surface support the first of these interpretations.  The early onset of surface motions also supports the first of these interpretations. 

Calculations of near-surface explosions \cite{Thomsen_1980} suggest that a steady-state flow field generated by Maxwell's Z-model can characterize the cratering flow field at any one time. We find this also to be true beyond the crater radius as shown in Figure \ref{fig:ray} subplot (b). When velocity peaks as shown in subplot (b), we compute the standard deviation of velocity angles subtracted by corresponding angles of the streamlines and find that it is 17${^\circ}$.  

Ray angles on the surface increase as a function of time before the velocity peaks and then slowly decrease.  A time dependent Maxwell's Z-model with $Z$ increasing with time (see \cite{Thomsen_1980}) might qualitatively match the temporal variation of our ray directions   

Unfortunately, two accelerometers in the SR5-ray experiment,  at $(R,z)$ = (5,0) and (5,-2), were temporarily saturated during the acceleration peak.
At these two positions, 
the angle $\vartheta$ is an overestimate while the accelerometers are saturated. 

\section{Discussion}
\label{sec:discussion}

\subsection{Comparing velocities outside the crater to estimated ejecta velocities}

\begin{figure}
    \centering
     \if \ispreprint0
    \includegraphics[scale = 0.8]{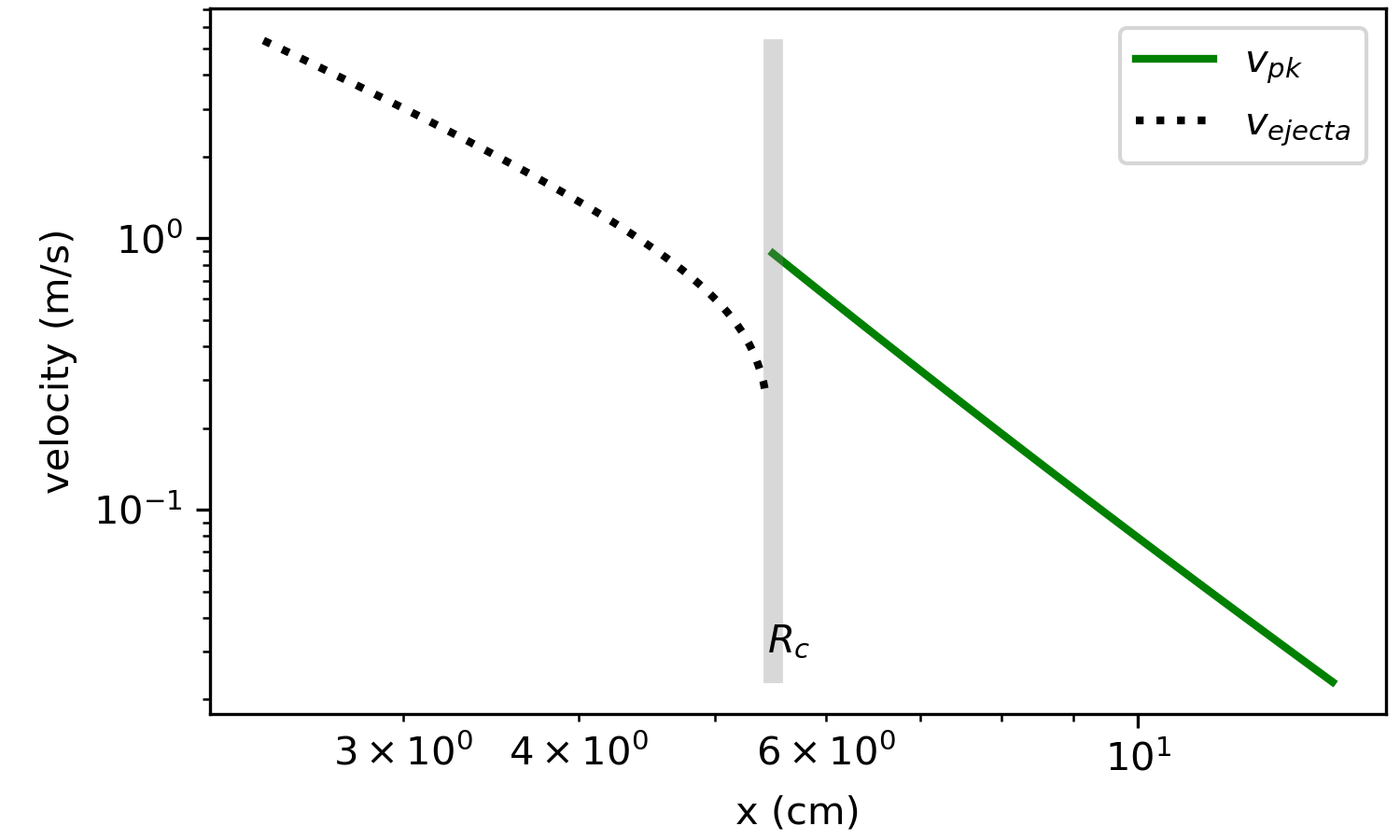}
    \else
    \includegraphics[width=3.2 truein, trim = 10 0 0 0 ,clip]{HH.png}
    \fi 
    \caption{Comparison of predicted ejecta velocities inside the crater radius and surface velocities outside the crater radius. Ejecta velocity as a function of launch position, derived by \citet{housen11} and given in equation \ref{eqn:vej}, is plotted with a black dotted line. The the solid green line is the peak velocity seen on the surface as a function of distance from the crater center.  We plot velocity magnitude using the power law fits given in Table \ref{tab:exp_fits}. The gray bar in the center indicates the crater radius.}
    \label{fig:vHH}
\end{figure}

The velocity vectors shown in Figure \ref{fig:ray} suggest that surface motions are a continuation of the crater excavation flow.  Are the surface velocities that we have measured outside the crater radius related to ejecta velocities?  We have not measured the velocity of ejecta particles in our experiments, however, we can compare the velocity of surface particles outside the crater to those predicted via scaling arguments for ejecta launched from within the crater. 

\citet{housen11} derived scaling laws for the properties of ejecta consistent with laboratory measurements. They showed that the ejecta velocity as a function of launch position $x$, the distance from the site of impact, is well described by their equation 14, which we rewrite  
\begin{equation}
    v_{ejecta}(x) = v_{imp} C_1 \left[ \frac{x}{R_p} \pi_4^\nu \right]^{-\frac{1}{\mu}}
\left(1 - \frac{x}{R_c} \right)^p. \label{eqn:vej}
\end{equation}
Our crater radius $R_c$ is the radius of the crater rim so is about 1.3 times the variable $R$ they use in their equation 14.   This equation predicts that ejecta has a velocity of zero at $R_c$, so only material within $R_c$ is ejected.  
Here $v_{imp}$ is the impact velocity, $R_p$ is the projectile radius and $\pi_4$ is the ratio of substrate to projectile density.  \citet{housen11} gave constant $C_1$ and exponents $\nu, \mu, p$
consistent with different experiments in their Table 3.
For these coefficients, we choose the lowest impact velocity experiment into a granular system, that labeled column C6 in their Table 3, which has coefficients $C_1 = 1$, $\nu=0.4$, $\mu  = 0.45$ and $p = 0.3$. 
Using these coefficients, $v_{ejecta(x)}$ from equation \ref{eqn:vej} is plotted in Figure \ref{fig:vHH} with a black dotted line.  The ejecta velocity drops to zero at the crater radius, as assumed by their model. 

In section \ref{sec:peak} we fit power-laws to the peak velocity of surface particles outside the crater radius.  We denote the distance from impact as $x$, so as to be consistent with equation \ref{eqn:vej}. Using the coefficients describing the power law fits listed in Table \ref{tab:power_fits}, we compute the peak velocity $v_{pk}(x) = \sqrt{v_{r,pk}^2 +v_{z,pk}^2}$. 
We plot $v_{pk}(x)$ as a solid green line in Figure \ref{fig:vHH}, but only outside the crater radius. 

Except for the region near the crater radius, where ejecta are assumed to have zero velocity,  Figure \ref{fig:vHH} shows that the predicted ejecta velocities could be a continuation of the velocities we see on the surface outside the crater radius.  
Accelerations outside the crater radius, shown in Figure \ref{fig:skin_delay} are above 1g, so vertical motions outside the crater on the surface are consistent with ejection being part of the uplift. 
 Figure 9a by \citet{Celik_2022} also shows that ejecta velocities extend beyond the crater radius in their DEM simulations.
Scaling laws for the properties of ejecta might be smoothly extended to estimate the extent of surface particle motions outside the crater radius.

\subsection{Comparing uplift and dilation to those estimated for impact craters}

A compilation of impacts in the gravity scaling regime in Figure 3 by \citet{Celik_2022} indicates that scaling laws for crater size in the gravity regime is remarkably consistent for impact velocities ranging from $\sim$ 1 m/s to 10 km/s. \citet{Cintala79} found that the ratio $h_{rim}/R_c$, where $h_{rim}$ is the crater rim height, is fairly constant among craters on different planets and so is approximately independent of gravity and impact energy.  
Structural uplift is the primary mechanism of rim development in both simple and complex lunar craters and complex Martian craters with structural uplift height exceeding the depth of the ejecta blanket by about a factor of 4 \citep{Sharpton_2014,Sturm_2016}.
\citet{Sharpton_2014} estimates that the crater rim structural uplift, $h_u \approx 0.8 \times h_{rim}$, with the other 20\% of the rim height profile resulting from ejecta. 
In our experiments, the uplift is directly related to the final vertical displacement at the crater radius; $h_{u} = \delta_{z,f}(R_c)$. 
We compare this ratio in our impact experiments to those measured for other impact craters. 


In our experiments we found that the crater rim uplift height was $\approx$2.5 mm for both our power law and exponential fits listed in Tables \ref{tab:power_fits} and  \ref{tab:exp_fits}. Using these final vertical displacements, we compute the ratio
\begin{equation}
    \mathcal{K}_z \equiv \frac{\delta_{z,f}(R_c)}{R_c} = \frac{h_u}{R_c}.
\end{equation}
For our experiments 
\begin{equation}
    \mathcal{K}_z \approx 0.05.
\end{equation}

If we use the SCI crater rim height at $R_{rim}$ in Figure 4 by \citet{Honda_2021} and assume that 20\% is ejecta, the rim displacement $\delta_{z,f}\approx 0.80 \times 0.47 = 0.376$ m, and $\mathcal{K}_z \approx 0.04\pm 0.01$. We estimate an error of 0.01 to take into account uncertainty in the fraction of ejecta.
The ratio $\mathcal{K}_z$  for the SCI impact is similar to that measured in our experiment.
Both craters are in the gravity regime, however the crater radii, impact velocity and gravity differ. 

\begin{figure}[htbp]
    \centering
    \includegraphics[scale = 0.55]{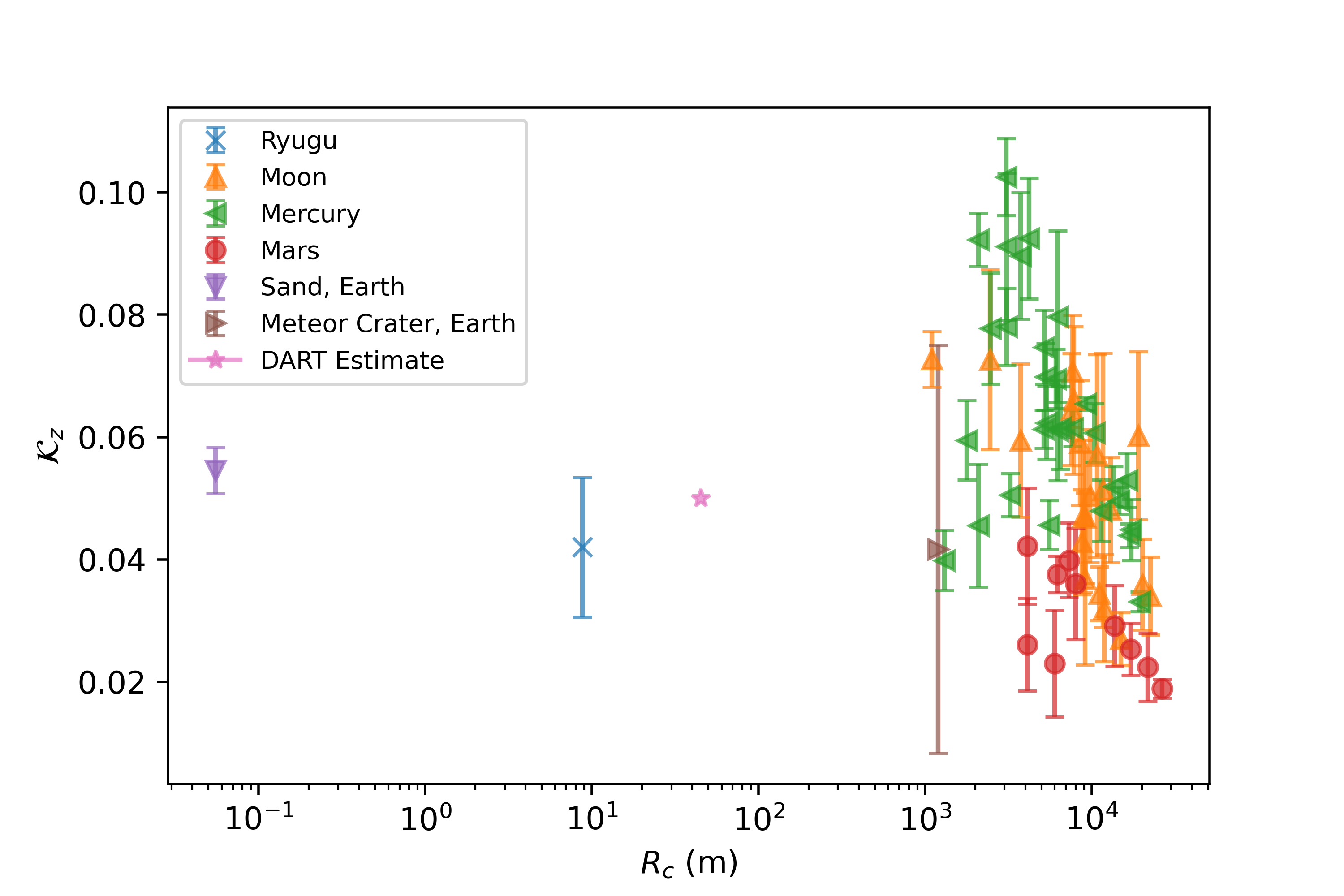}
    \caption{Plot of the ratio of rim uplift to crater radius, $\mathcal{K}_z$, versus crater radius for impacts on several different bodies. Lunar data is from \citet{Sharpton_2014}, Martian data is from \citet{Sturm_2016}, Mercury data from \citet{Cintala79}, Ryugu data from \citet{Honda_2021}, sand data from our low velocity experiments, and the Meteor Crater data is from \citet{Poelchau_2009}. We also plot an estimate for the DART mission impact, assuming the impact crater rim has a similar ratio to the SCI impact crater. Our experiments and the SCI impact, both of which are in the gravity regime, have $\mathcal{K}_z$ ratios that agree with craters several orders of magnitude greater in size.}
    \label{fig:gravity_invariant}
\end{figure}

The ratio of structural uplift height to crater radius for simple lunar craters is about $\mathcal{K}_z \sim 0.07$  \citep{Sharpton_2014}. 
For Martian craters with $R_c \approx 20$ km, $\mathcal{K}_z \sim 0.025$ \citet{Sturm_2016}. 
Figure \ref{fig:gravity_invariant} shows a compilation of structural uplift comparisons based on studies of craters on the Moon \citep{Sharpton_2014}, Mercury \citep{Cintala79}, Mars \citep{Sturm_2016}, Earth (this paper and from meteor crater data \citep{Poelchau_2009}), and Ryugu \citep{Honda_2021}.
The ratio $\mathcal{K}_z$ for structural uplift on these bodies ranges from about 0.02 to 0.1. 
Our experiment is the only low velocity experiment in this compilation, nevertheless the ratio estimated in our experiment lies approximately at the median of the other measurements.
Our experimental value for the ratio of rim uplift to crater radius supports the proposal \citep{Cintala79} that this ratio is independent of impact energy, gravity, and surface strength, for impact craters that are in the gravity regime. 

In 2026, ESA's Hera probe will image the impact crater from the DART mission. The Hera probe will have an altimeter resolution of 0.5 m \citep{altimeter}. If we assume the DART impact is in the gravity regime and follows a similar $\mathcal{K}_z = 0.05$ ratio as our experiments and the SCI impact, the crater radius $R_c$ has to be $\geq 10$ m for the rim uplift $h_u$ to be detectable. \citet{Cheng_2020} estimated a wide range of crater radii for the DART impact using scaling laws \citep{housen11}, where $R_c \sim$ 4 to 45 m. These estimates depend on substrate material properties and the smaller crater radii correspond to impacts in the strength dominated regime. The largest estimated crater radius was 45 m, from impacts into a sand-like substrate with a porosity of 35\% and was in the gravity regime. The DART estimate in our Figure \ref{fig:gravity_invariant} uses the $R_c = 45$ m estimate and $\mathcal{K}_z = 0.05$.  If the crater radius is 45 m and $\mathcal{K}_z = 0.05$, then the uplift $h_u=2$ m,  which would be resolved by the Hera mission. 

\subsection{Scaling radial surface displacement to crater radius}

How does a similar ratio, 
\begin{equation} 
\mathcal{K}_r = \frac{\delta_{r,f}(R_c)}{R_c}, \label{eqn:Kr}
\end{equation}
for final radial displacements compare to those measured in other settings?  The final radial displacement correspond to dilation or widening of the crater, rather than uplift. 
To answer this question, we compare radial displacements measured in our experiments to those measured post-impact from analysis of the artificial, high speed impact on the asteroid Ryugu \citep{Arakawa_2020,Honda_2021}.  

In our experiments we found that the final radial displacement at one crater radius $\delta_{r,f}(R_c) \approx 6.6$ mm (using the power-law fits listed in Table \ref{tab:power_fits}). Using these final radial displacements we compute $\mathcal{K}_r \approx 0.12$. The value $\delta_{r,f}$ from our exponential fits gives a lower value  $\mathcal{K}_r\approx0.10$ so the uncertainty in our estimate is about 20\%. 
For the SCI crater, we estimate the final displacement of boulders to be around 1 meter, which is within the scatter of Figure 6 by \citet{Honda_2021}.  This with a crater radius of 8.8 m gives $\mathcal{K}_r\approx0.11$ for the SCI impact crater on Ryugu. Since these $\mathcal{K}_r$ ratios are approximately equivalent, we postulate that $\mathcal{K}_r$ is also independent of gravity. Applying this ratio to the DART impact and using $R_c = 45$ m by \citet{Cheng_2020}, we would predict a final displacement $\delta_{r,f}(R_c) = 5.4$ m.

Since LICIACube will take approximately 5 images before and after the DART impact \citep{Dotto_2021}, it is unlikely we will be able to observe the peak velocity of the surface pulse.  Out of curiosity, we estimate the peak velocities for the SCI and DART impacts.  If we assume that the surface pulse travels the distance particles are displaced at the crater radius during the crater excavation or formation time, $\tau_{ex}$, the peak surface velocity near the crater radius can be estimated with the ratio
\begin{equation}
    v_s \sim \frac{\delta_{r,f}(R_c)}{\tau_{ex}}.
    \label{eqn:vest1}
\end{equation}
Using our definition for $\mathcal{K}_r$ in equation \ref{eqn:Kr}, and the crater formation time in the gravity regime estimated by \cite{Housen_1983}, $\tau_{ex} \sim \sqrt{R_c/g}$,  
\begin{equation}
    v_s \sim \mathcal{K}_r \frac{R_c}{\tau_{ex}}
    \sim \mathcal{K}_r \sqrt{g R_c}.
    \label{eqn:vest2}
\end{equation}

Using $\mathcal{K}_r = 0.1$ and the crater radius
for our experiments,
Equation \ref{eqn:vest2} gives $v_s \sim 7$ cm/s, 
which is lower, but within a factor of a few  of the  magnitude of the peak surface velocity 
that we observed outside the crater radius.  For the SCI impact, with $g \approx 1.2 \times 10^{-4}$ \citep{Arakawa_2020} equation \ref{eqn:vest2} gives an 
 estimate of $v_s \approx 0.3$ cm/s.
For the DART impact, with crater radius of 45 m
and $g \approx 5 \times 10^{-5}$ \citep{Naidu_2020}
equation \ref{eqn:vest2} gives a velocity of only 
$v_s \sim 0.5 $ cm/s.
The surface displacement is only 1 m after about 3 minutes, making it difficult for LICIACube to detect, even at its close approach to Dimorphos where the resolution of its PL1 camera is about a meter.

\section{Summary}

In this study we have tracked motions of surface particles that are induced by a low velocity normal impact in a granular medium using high speed videos.
We have succeeded in detecting the surface particle motions with a combination of methods, tracking individual fluorescent particles laying on the surface and with a cross correlation method.  
For impacts of marble sized projectiles into sand with velocity a few m/s, displacements of surface particles are small, only about a few mm just outside the crater radius. 

Both plastic, permanent deformations and elastic, rebounding deformations are observed.
Particle motions begin a few ms after the impact and before launch of most of the ejecta curtain. 
The peak displacement and final displacement are sensitive to distance from the site of impact and decay with a power law form $\propto r^{-\beta}$ with index $\beta \sim -4$. The exponent is steeper than the -2 predicted for the pulse displacement peak using a diffusive pulse propagation model for the propagation of a subsurface seismic pulse \citep{Quillen_2022}.

The duration of time during which the surface particles are moving is broad, extending up to 60 ms near the crater radius.  This duration  exceeds the widths of subsurface seismic pulses (a few ms) and suggests that the surface motions are associated with crater excavation which takes place on a similar longer time.

We find that motions of particles on the surface differ from those a few cm below the surface.   While the amplitude of peak velocity is similar, the velocity pulses are longer duration on the surface than a few cm below.  The onset of motion on the surface is delayed compared to motions seen below surface.  Peak accelerations are lower on the surface than in the sub-surface seismic pulse.

We examine the ray angles for the velocity vectors as a function of position and time. 
When the velocity peaks, the velocity field resembles the flow field predicted for crater excavation
for a Maxwell's Z model \citep{Maxwell_1977} with a $Z$ value of about 3.5. This suggests that the flow field outside the crater excavation region is a continuation of that during crater excavation and within the transient crater. Assuming Maxwell's Z-model can represent the flow field beyond the crater excavation region at any time, we find that Z increases with time, supporting \citet{Thomsen_1980}. 

The ratio of final particle vertical displacement to crater radius in our experiments is similar to the ratio of structure rim uplift to crater radius measured for other craters that lie in the gravity regime.   This supports the proposal
that this ratio is nearly independent of many crater scaling parameters \citep{Cintala79}.

\appendix
\if \ispreprint0
\section{Supplementary Videos}

Below are images taken from the supplementary videos included with the manuscript.


\begin{figure*}[] 
    \centering
    \begin{subfigure}[t]{0.9\textwidth}
        \centering
        \includegraphics[width=\textwidth]{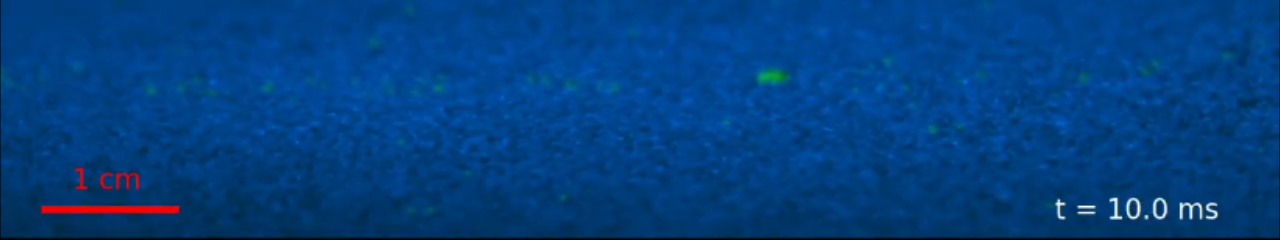}
        \caption{Frame from supplemental video PTVs-01}
        \label{fig:side_screenshot}
    \end{subfigure}
    \begin{subfigure}[t]{0.9\textwidth}
        \centering
        \includegraphics[width=\textwidth]{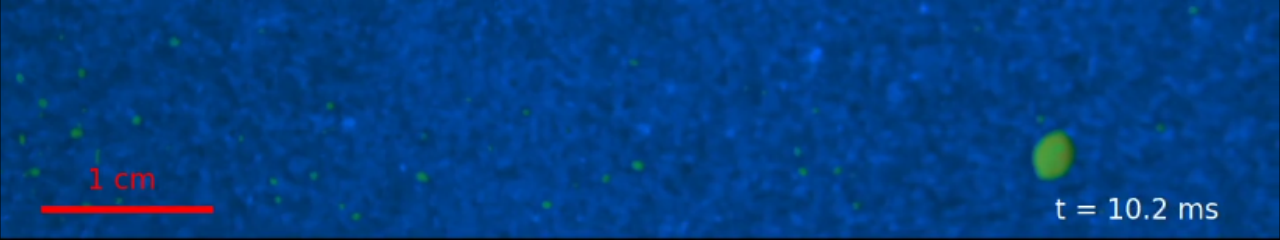}
        \caption{Frame from supplemental video PTVa-01.}
        \label{fig:above_screenshot}
    \end{subfigure}
    \begin{subfigure}[t]{0.9\textwidth}
        \centering
       \includegraphics[width=\textwidth]{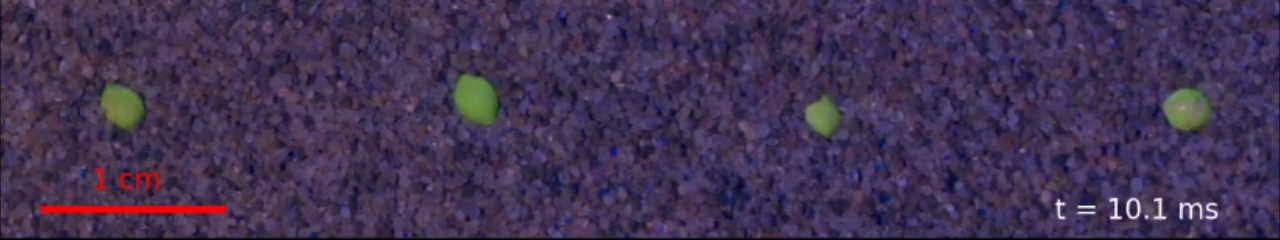}
        \caption{Frame from supplemental video CC-01}
        \label{fig:cross_corr_screenshot}
    \end{subfigure}
    \begin{subfigure}[t]{0.9\textwidth}
        \centering
        \includegraphics[width=\textwidth]{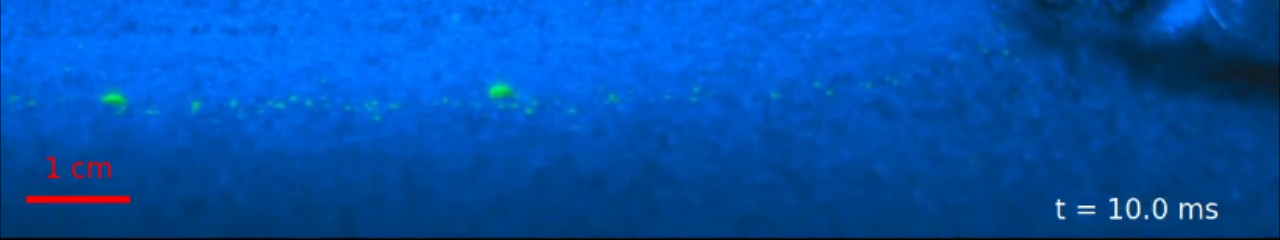}
        \caption{Frame from supplemental video PTVs-compare}
        \label{fig:compare_screenshot}
    \end{subfigure}
    \begin{subfigure}[t]{0.5\textwidth}
        \centering
        \includegraphics[width=\textwidth,scale=0.5,clip]{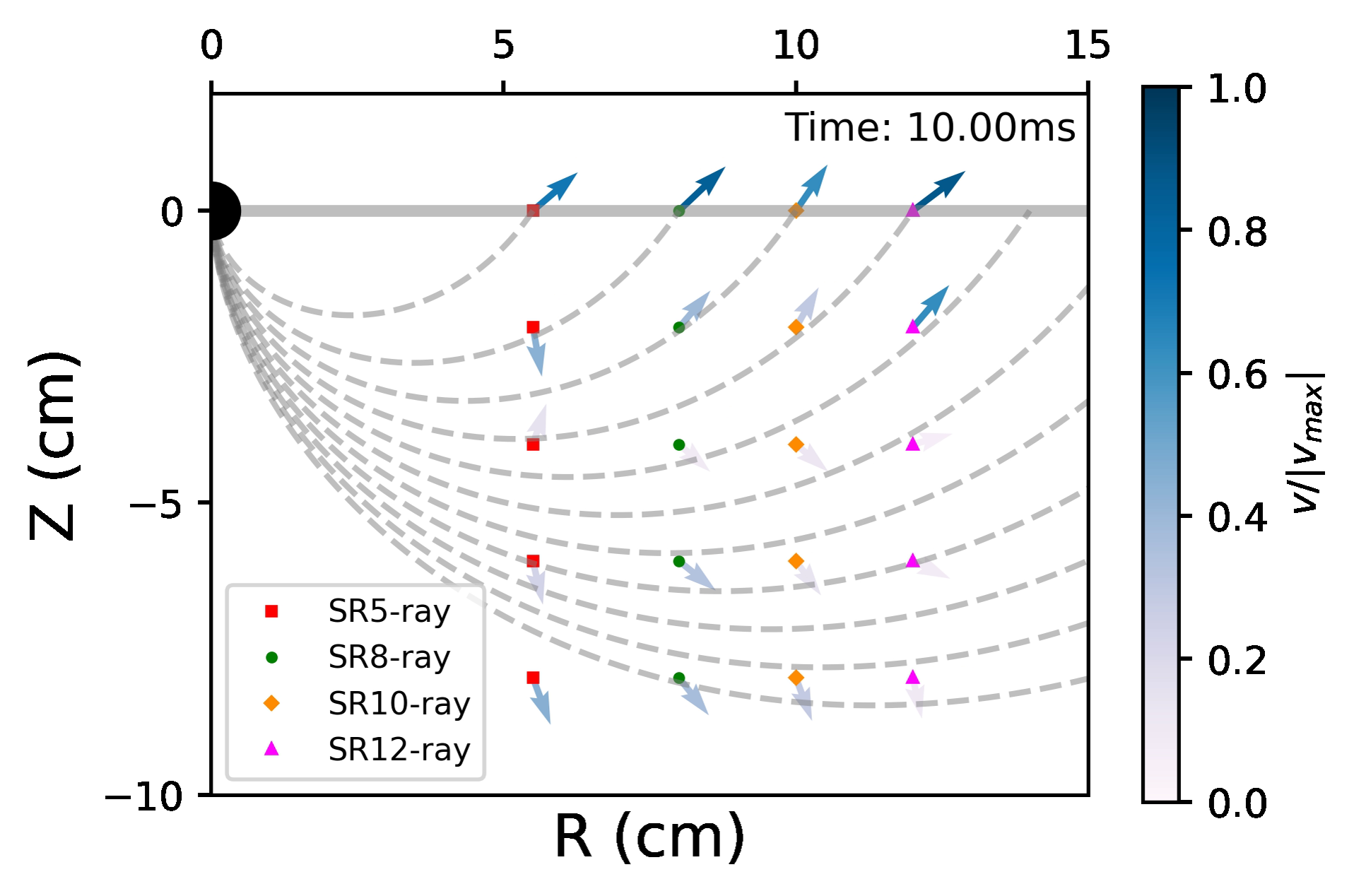}
        \caption{Frame from supplemental video `ray\_angle'}
        \label{fig:ray_angle_screenshot}
    \end{subfigure}
    \caption{Individual frames of supplemental videos (from top to bottom) PTVs-01, PTVa-01, CC-01, PTVs-compare, and 'ray\_angle'. Details regarding these videos are listed in Table \ref{tab:exp_videos_list}. See Section \ref{sec:ray} for details about video 'ray\_angle'.}
    \label{fig:supplementary_video_screenshots}
\end{figure*}
\fi

\vskip 2 truein
{Acknowledgements:}
This material is based upon work supported in part by NASA grant 80NSSC21K0143, and the Schwartz Discover Grant to M.N. at the University of Rochester.

Scripts used to make the figures in this document are available at  \url{https://github.com/URGranularLab/Surface_Pulse}

\bibliographystyle{elsarticle-harv}
\bibliography{refs_atten}

\begin{thebibliography}{44}
\expandafter\ifx\csname natexlab\endcsname\relax\def\natexlab#1{#1}\fi
\providecommand{\url}[1]{\texttt{#1}}
\providecommand{\href}[2]{#2}
\providecommand{\path}[1]{#1}
\providecommand{\DOIprefix}{doi:}
\providecommand{\ArXivprefix}{arXiv:}
\providecommand{\URLprefix}{URL: }
\providecommand{\Pubmedprefix}{pmid:}
\providecommand{\doi}[1]{\href{http://dx.doi.org/#1}{\path{#1}}}
\providecommand{\Pubmed}[1]{\href{pmid:#1}{\path{#1}}}
\providecommand{\bibinfo}[2]{#2}
\ifx\xfnm\relax \def\xfnm[#1]{\unskip,\space#1}\fi
\bibitem[{Allan et~al.()Allan, Caswell, Keim and van~der Wel}]{trackpy}
\bibinfo{author}{Allan, D.}, \bibinfo{author}{Caswell, T.},
  \bibinfo{author}{Keim, N.}, \bibinfo{author}{van~der Wel, C.}, .
\newblock \bibinfo{title}{Trackpy v0.3.2}.
\newblock \URLprefix \url{https://doi.org/10.5281/zenodo.60550}.
\bibitem[{Arakawa et~al.(2020)Arakawa, Saiki, Wada, Ogawa, Kadono, Shirai,
  Sawada, Ishibashi, Honda, Sakatani, Iijima, Okamoto, Yano, Takagi, Hayakawa,
  Michel, Jutzi, Shimaki, Kimura, Mimasu, Toda, Imamura, Nakazawa, Hayakawa,
  Sugita, Morota, Kameda, Tatsumi, Cho, Yoshioka, Yokota, Matsuoka, Yamada,
  Kouyama, Honda, Tsuda, Watanabe, Yoshikawa, Tanaka, Terui, Kikuchi,
  Yamaguchi, Ogawa, Ono, Yoshikawa, Takahashi, Takei, Fujii, Takeuchi,
  Yamamoto, Okada, Hirose, Hosoda, Mori, Shimada, Soldini, Tsukizaki, Iwata,
  Ozaki, Abe, Namiki, Kitazato, Tachibana, Ikeda, Hirata, Hirata, Noguchi and
  Miura}]{Arakawa_2020}
\bibinfo{author}{Arakawa, M.}, \bibinfo{author}{Saiki, T.},
  \bibinfo{author}{Wada, K.}, \bibinfo{author}{Ogawa, K.},
  \bibinfo{author}{Kadono, T.}, \bibinfo{author}{Shirai, K.},
  \bibinfo{author}{Sawada, H.}, \bibinfo{author}{Ishibashi, K.},
  \bibinfo{author}{Honda, R.}, \bibinfo{author}{Sakatani, N.},
  \bibinfo{author}{Iijima, Y.}, \bibinfo{author}{Okamoto, C.},
  \bibinfo{author}{Yano, H.}, \bibinfo{author}{Takagi, Y.},
  \bibinfo{author}{Hayakawa, M.}, \bibinfo{author}{Michel, P.},
  \bibinfo{author}{Jutzi, M.}, \bibinfo{author}{Shimaki, Y.},
  \bibinfo{author}{Kimura, S.}, \bibinfo{author}{Mimasu, Y.},
  \bibinfo{author}{Toda, T.}, \bibinfo{author}{Imamura, H.},
  \bibinfo{author}{Nakazawa, S.}, \bibinfo{author}{Hayakawa, H.},
  \bibinfo{author}{Sugita, S.}, \bibinfo{author}{Morota, T.},
  \bibinfo{author}{Kameda, S.}, \bibinfo{author}{Tatsumi, E.},
  \bibinfo{author}{Cho, Y.}, \bibinfo{author}{Yoshioka, K.},
  \bibinfo{author}{Yokota, Y.}, \bibinfo{author}{Matsuoka, M.},
  \bibinfo{author}{Yamada, M.}, \bibinfo{author}{Kouyama, T.},
  \bibinfo{author}{Honda, C.}, \bibinfo{author}{Tsuda, Y.},
  \bibinfo{author}{Watanabe, S.}, \bibinfo{author}{Yoshikawa, M.},
  \bibinfo{author}{Tanaka, S.}, \bibinfo{author}{Terui, F.},
  \bibinfo{author}{Kikuchi, S.}, \bibinfo{author}{Yamaguchi, T.},
  \bibinfo{author}{Ogawa, N.}, \bibinfo{author}{Ono, G.},
  \bibinfo{author}{Yoshikawa, K.}, \bibinfo{author}{Takahashi, T.},
  \bibinfo{author}{Takei, Y.}, \bibinfo{author}{Fujii, A.},
  \bibinfo{author}{Takeuchi, H.}, \bibinfo{author}{Yamamoto, Y.},
  \bibinfo{author}{Okada, T.}, \bibinfo{author}{Hirose, C.},
  \bibinfo{author}{Hosoda, S.}, \bibinfo{author}{Mori, O.},
  \bibinfo{author}{Shimada, T.}, \bibinfo{author}{Soldini, S.},
  \bibinfo{author}{Tsukizaki, R.}, \bibinfo{author}{Iwata, T.},
  \bibinfo{author}{Ozaki, M.}, \bibinfo{author}{Abe, M.},
  \bibinfo{author}{Namiki, N.}, \bibinfo{author}{Kitazato, K.},
  \bibinfo{author}{Tachibana, S.}, \bibinfo{author}{Ikeda, H.},
  \bibinfo{author}{Hirata, N.}, \bibinfo{author}{Hirata, N.},
  \bibinfo{author}{Noguchi, R.}, \bibinfo{author}{Miura, A.},
  \bibinfo{year}{2020}.
\newblock \bibinfo{title}{An artificial impact on the asteroid (162173) {Ryugu}
  formed a crater in the gravity-dominated regime}.
\newblock \bibinfo{journal}{Science} \bibinfo{volume}{368},
  \bibinfo{pages}{67--71}.
\newblock \URLprefix \url{https://doi.org/10.1126%2Fscience.aaz1701},
  \DOIprefix\doi{10.1126/science.aaz1701}.
\bibitem[{Austin et~al.(1981)Austin, Thomsen, Ruhl, Orphal, Borden, Larson and
  Schultz}]{Austin_1981}
\bibinfo{author}{Austin, M.G.}, \bibinfo{author}{Thomsen, J.M.},
  \bibinfo{author}{Ruhl, S.F.}, \bibinfo{author}{Orphal, D.L.},
  \bibinfo{author}{Borden, W.F.}, \bibinfo{author}{Larson, S.A.},
  \bibinfo{author}{Schultz, P.H.}, \bibinfo{year}{1981}.
\newblock \bibinfo{title}{Z-model analysis of impact cratering: An overview},
  in: \bibinfo{editor}{Schultz, P.H.}, \bibinfo{editor}{Merrill, R.B.} (Eds.),
  \bibinfo{booktitle}{Multi-ring Basins, Proceedings of Lunar and Planetary
  Science}, \bibinfo{publisher}{Pergamon Press, New York.}. pp.
  \bibinfo{pages}{197--205}.
\bibitem[{\c{C}elic et~al.(2022)\c{C}elic, Ballouz, Scheeres and
  Kawakatsu}]{Celik_2022}
\bibinfo{author}{\c{C}elic, O.}, \bibinfo{author}{Ballouz, R.L.},
  \bibinfo{author}{Scheeres, D.L.}, \bibinfo{author}{Kawakatsu, Y.},
  \bibinfo{year}{2022}.
\newblock \bibinfo{title}{A numerical simulation approach to the crater-scaling
  relationships in low-speed impacts under microgravity}.
\newblock \bibinfo{journal}{Icarus} \bibinfo{volume}{377},
  \bibinfo{pages}{114882}.
\bibitem[{Cheng et~al.(2018)Cheng, Rivkin, Michel, Atchison, Barnouin, Benner,
  Chabot, Ernst, Fahnestock, Kueppers, Pravec, Rainey, Richardson, Stickle and
  Thomas}]{Cheng_2018_DART}
\bibinfo{author}{Cheng, A.F.}, \bibinfo{author}{Rivkin, A.S.},
  \bibinfo{author}{Michel, P.}, \bibinfo{author}{Atchison, J.},
  \bibinfo{author}{Barnouin, O.}, \bibinfo{author}{Benner, L.},
  \bibinfo{author}{Chabot, N.L.}, \bibinfo{author}{Ernst, C.},
  \bibinfo{author}{Fahnestock, E.G.}, \bibinfo{author}{Kueppers, M.},
  \bibinfo{author}{Pravec, P.}, \bibinfo{author}{Rainey, E.},
  \bibinfo{author}{Richardson, D.C.}, \bibinfo{author}{Stickle, A.M.},
  \bibinfo{author}{Thomas, C.}, \bibinfo{year}{2018}.
\newblock \bibinfo{title}{{AIDA DART} asteroid deflection test: Planetary
  defense and science objectives}.
\newblock \bibinfo{journal}{Planetary and Space Science} \bibinfo{volume}{157},
  \bibinfo{pages}{104--115}.
\newblock \DOIprefix\doi{https://doi.org/10.1016/j.pss.2018.02.015}.
\bibitem[{Cheng et~al.(2020)Cheng, Stickle, Fahnestock, Dotto, Corte, Chabot
  and Rivkin}]{Cheng_2020}
\bibinfo{author}{Cheng, A.F.}, \bibinfo{author}{Stickle, A.M.},
  \bibinfo{author}{Fahnestock, E.G.}, \bibinfo{author}{Dotto, E.},
  \bibinfo{author}{Corte, V.D.}, \bibinfo{author}{Chabot, N.L.},
  \bibinfo{author}{Rivkin, A.S.}, \bibinfo{year}{2020}.
\newblock \bibinfo{title}{{DART} mission determination of momentum transfer:
  Model of ejecta plume observations}.
\newblock \bibinfo{journal}{Icarus} \bibinfo{volume}{352},
  \bibinfo{pages}{113989}.
\newblock \DOIprefix\doi{10.1016/j.icarus.2020.113989}.
\bibitem[{Cintala(1979)}]{Cintala79}
\bibinfo{author}{Cintala, M.J.}, \bibinfo{year}{1979}.
\newblock \bibinfo{title}{Mercurian crater rim heights and some interplanetary
  comparisons}, in: \bibinfo{booktitle}{Lunar and Planetary Science Conference,
  10th, Houston, Tex.}, pp. \bibinfo{pages}{2635--2650}.
\newblock \URLprefix
  \url{https://articles.adsabs.harvard.edu//full/1979LPSC...10.2635C/0002635.000.html}.
\bibitem[{Crocker and Grier(1996)}]{crocker96}
\bibinfo{author}{Crocker, J.C.}, \bibinfo{author}{Grier, D.G.},
  \bibinfo{year}{1996}.
\newblock \bibinfo{title}{Methods of digital video microscopy for colloidal
  studies}.
\newblock \bibinfo{journal}{Journal of Colloid Interface Science}
  \bibinfo{volume}{179}, \bibinfo{pages}{298--310}.
\bibitem[{Croft(1980)}]{Croft_1980}
\bibinfo{author}{Croft, S.K.}, \bibinfo{year}{1980}.
\newblock \bibinfo{title}{Cratering flow fields: Implications for the
  excavation and stranient expansion stages of crater formation}, in:
  \bibinfo{editor}{Institute, L.S.} (Ed.), \bibinfo{booktitle}{Proceedings of
  the IIIrd Intern. Conf. on Powders \& Grains (Balkema, Rotterdam) of the
  Lunar Planet. Sci. Conf. 11th}, \bibinfo{organization}{American Geophysical
  Union}. \bibinfo{publisher}{New York : Pergamon Press}. pp.
  \bibinfo{pages}{2347--2378}.
\bibitem[{Dotto et~al.(2021)Dotto, {Della Corte}, Amoroso, Bertini, Brucato,
  Capannolo, Cotugno, Cremonese, {Di Tana}, Gai, Ieva, Impresario, Ivanovski,
  Lavagna, Lucchetti, {Mazzotta Epifani}, Meneghin, Miglioretti, Modenini,
  Pajola, Palumbo, Perna, Pirrotta, Poggiali, Rossi, Simioni, Simonetti,
  Tortora, Zannoni, Zanotti, Zinzi, Cheng, Rivkin, Adams, Reynolds and
  Fretz}]{Dotto_2021}
\bibinfo{author}{Dotto, E.}, \bibinfo{author}{{Della Corte}, V.},
  \bibinfo{author}{Amoroso, M.}, \bibinfo{author}{Bertini, I.},
  \bibinfo{author}{Brucato, J.}, \bibinfo{author}{Capannolo, A.},
  \bibinfo{author}{Cotugno, B.}, \bibinfo{author}{Cremonese, G.},
  \bibinfo{author}{{Di Tana}, V.}, \bibinfo{author}{Gai, I.},
  \bibinfo{author}{Ieva, S.}, \bibinfo{author}{Impresario, G.},
  \bibinfo{author}{Ivanovski, S.}, \bibinfo{author}{Lavagna, M.},
  \bibinfo{author}{Lucchetti, A.}, \bibinfo{author}{{Mazzotta Epifani}, E.},
  \bibinfo{author}{Meneghin, A.}, \bibinfo{author}{Miglioretti, F.},
  \bibinfo{author}{Modenini, D.}, \bibinfo{author}{Pajola, M.},
  \bibinfo{author}{Palumbo, P.}, \bibinfo{author}{Perna, D.},
  \bibinfo{author}{Pirrotta, S.}, \bibinfo{author}{Poggiali, G.},
  \bibinfo{author}{Rossi, A.}, \bibinfo{author}{Simioni, E.},
  \bibinfo{author}{Simonetti, S.}, \bibinfo{author}{Tortora, P.},
  \bibinfo{author}{Zannoni, M.}, \bibinfo{author}{Zanotti, G.},
  \bibinfo{author}{Zinzi, A.}, \bibinfo{author}{Cheng, A.},
  \bibinfo{author}{Rivkin, A.}, \bibinfo{author}{Adams, E.},
  \bibinfo{author}{Reynolds, E.}, \bibinfo{author}{Fretz, K.},
  \bibinfo{year}{2021}.
\newblock \bibinfo{title}{{LICIACube} - the {Light} {Italian} {Cubesat} for
  imaging of asteroids in support of the {NASA DART} mission towards asteroid
  (65803) {Didymos}}.
\newblock \bibinfo{journal}{Planetary and Space Science} \bibinfo{volume}{199},
  \bibinfo{pages}{105185}.
\newblock \DOIprefix\doi{https://doi.org/10.1016/j.pss.2021.105185}.
\bibitem[{Duffy and Mindlin(1957)}]{Duffy_1957}
\bibinfo{author}{Duffy, J.}, \bibinfo{author}{Mindlin, R.},
  \bibinfo{year}{1957}.
\newblock \bibinfo{title}{Stress-strain relations and vibrations of a granular
  medium}.
\newblock \bibinfo{journal}{J. Appl. Mech.} \bibinfo{volume}{24},
  \bibinfo{pages}{585--593}.
\bibitem[{Holsapple(1993)}]{holsapple93}
\bibinfo{author}{Holsapple, K.A.}, \bibinfo{year}{1993}.
\newblock \bibinfo{title}{The scaling of impact processes in planetary
  sciences}.
\newblock \bibinfo{journal}{Annual Review of Earth and Planetary Sciences}
  \bibinfo{volume}{21}, \bibinfo{pages}{333--373}.
\bibitem[{Honda et~al.(2021)Honda, Arakawa, Shimaki, Shirai, Yokota, Kadono,
  Wada, Ogawa, Ishibashi, Sakatani, Nakazawa, Yasui, Morota, Kameda, Tatsumi,
  Yamada, Kouyama, Cho, Matsuoka, Suzuki, Honda, Hayakawa, Yoshioka, Hirata,
  Hirata, Sawada, Sugita, Saiki, Imamura, Takagi, Yano, Okamoto, Tsuda and ichi
  Iijima}]{Honda_2021}
\bibinfo{author}{Honda, R.}, \bibinfo{author}{Arakawa, M.},
  \bibinfo{author}{Shimaki, Y.}, \bibinfo{author}{Shirai, K.},
  \bibinfo{author}{Yokota, Y.}, \bibinfo{author}{Kadono, T.},
  \bibinfo{author}{Wada, K.}, \bibinfo{author}{Ogawa, K.},
  \bibinfo{author}{Ishibashi, K.}, \bibinfo{author}{Sakatani, N.},
  \bibinfo{author}{Nakazawa, S.}, \bibinfo{author}{Yasui, M.},
  \bibinfo{author}{Morota, T.}, \bibinfo{author}{Kameda, S.},
  \bibinfo{author}{Tatsumi, E.}, \bibinfo{author}{Yamada, M.},
  \bibinfo{author}{Kouyama, T.}, \bibinfo{author}{Cho, Y.},
  \bibinfo{author}{Matsuoka, M.}, \bibinfo{author}{Suzuki, H.},
  \bibinfo{author}{Honda, C.}, \bibinfo{author}{Hayakawa, M.},
  \bibinfo{author}{Yoshioka, K.}, \bibinfo{author}{Hirata, N.},
  \bibinfo{author}{Hirata, N.}, \bibinfo{author}{Sawada, H.},
  \bibinfo{author}{Sugita, S.}, \bibinfo{author}{Saiki, T.},
  \bibinfo{author}{Imamura, H.}, \bibinfo{author}{Takagi, Y.},
  \bibinfo{author}{Yano, H.}, \bibinfo{author}{Okamoto, C.},
  \bibinfo{author}{Tsuda, Y.}, \bibinfo{author}{ichi Iijima, Y.},
  \bibinfo{year}{2021}.
\newblock \bibinfo{title}{Resurfacing processes on asteroid (162173) {Ryugu}
  caused by an artificial impact of {Hayabusa 2's} small carry-on impactor}.
\newblock \bibinfo{journal}{Icarus} \bibinfo{volume}{366},
  \bibinfo{pages}{114530}.
\newblock \DOIprefix\doi{https://doi.org/10.1016/j.icarus.2021.114530}.
\bibitem[{Housen and Holsapple(2011)}]{housen11}
\bibinfo{author}{Housen, K.R.}, \bibinfo{author}{Holsapple, K.A.},
  \bibinfo{year}{2011}.
\newblock \bibinfo{title}{Ejecta from impact craters}.
\newblock \bibinfo{journal}{Icarus} \bibinfo{volume}{211},
  \bibinfo{pages}{856--875}.
\bibitem[{Housen et~al.(1983)Housen, Schmidt and Holsapple}]{Housen_1983}
\bibinfo{author}{Housen, K.R.}, \bibinfo{author}{Schmidt, R.M.},
  \bibinfo{author}{Holsapple, K.A.}, \bibinfo{year}{1983}.
\newblock \bibinfo{title}{Crater ejecta scaling laws: Fundamental forms based
  on dimensional analysis}.
\newblock \bibinfo{journal}{J. Geophys. Res.} \bibinfo{volume}{88},
  \bibinfo{pages}{2485--2499}.
\bibitem[{Johnson et~al.(2000)Johnson, Makse, Gland and
  Schwartz}]{Johnson_2000}
\bibinfo{author}{Johnson, D.L.}, \bibinfo{author}{Makse, H.A.},
  \bibinfo{author}{Gland, N.}, \bibinfo{author}{Schwartz, L.},
  \bibinfo{year}{2000}.
\newblock \bibinfo{title}{Nonlinear elasticity of granular media}.
\newblock \bibinfo{journal}{Physica B} \bibinfo{volume}{279},
  \bibinfo{pages}{134--138}.
\bibitem[{Kurosawa(2019)}]{Kurosawa_2019}
\bibinfo{author}{Kurosawa, K.}, \bibinfo{year}{2019}.
\newblock \bibinfo{title}{Impact cratering mechanics: A forward approach to
  predicting ejecta velocity distribution and transient crater radii}.
\newblock \bibinfo{journal}{Icarus} \bibinfo{volume}{317},
  \bibinfo{pages}{135--147}.
\bibitem[{Liu and Nagel(1992)}]{Liu_1992}
\bibinfo{author}{Liu, C.H.}, \bibinfo{author}{Nagel, S.R.},
  \bibinfo{year}{1992}.
\newblock \bibinfo{title}{Sound in sand}.
\newblock \bibinfo{journal}{Phys. Rev. Lett.} \bibinfo{volume}{68},
  \bibinfo{pages}{2301--2304}.
\bibitem[{Matsue et~al.(2020)Matsue, Yasui, Arakawa and Hasegawa}]{Matsue_2020}
\bibinfo{author}{Matsue, K.}, \bibinfo{author}{Yasui, M.},
  \bibinfo{author}{Arakawa, M.}, \bibinfo{author}{Hasegawa, S.},
  \bibinfo{year}{2020}.
\newblock \bibinfo{title}{Measurements of seismic waves induced by
  high-velocity impacts: Implications for seismic shaking surrounding impact
  craters on asteroids}.
\newblock \bibinfo{journal}{Icarus} \bibinfo{volume}{338},
  \bibinfo{pages}{113520}.
\bibitem[{Maxwell(1977)}]{Maxwell_1977}
\bibinfo{author}{Maxwell, D.E.}, \bibinfo{year}{1977}.
\newblock \bibinfo{title}{Simple z model of cratering, ejection, and the
  overturned flap}, in: \bibinfo{editor}{Roddy, D.J.}, \bibinfo{editor}{Pepin,
  R.O.}, \bibinfo{editor}{Merrill, R.B.} (Eds.), \bibinfo{booktitle}{Planetary
  and terrestrial implications: proceedings of the Symposium on Planetary
  Cratering Mechanics, Flagstaff, Arizona, September 13-17, 1976},
  \bibinfo{publisher}{New York: Pergamon Press}. pp.
  \bibinfo{pages}{1003--1008}.
\bibitem[{Melosh(1985)}]{Melosh_1985}
\bibinfo{author}{Melosh, H.J.}, \bibinfo{year}{1985}.
\newblock \bibinfo{title}{Impact cratering mechanics: Relationship between the
  shock wave and excavation flow}.
\newblock \bibinfo{journal}{Icarus} \bibinfo{volume}{62},
  \bibinfo{pages}{339--343}.
\bibitem[{Melosh(1989)}]{melosh89}
\bibinfo{author}{Melosh, H.J.}, \bibinfo{year}{1989}.
\newblock \bibinfo{title}{Impact cratering: a geologic process}.
\newblock Oxford Monographs on Geology and Geophysics,
  \bibinfo{publisher}{Oxford University Press, Oxford England}.
\bibitem[{Melosh and Ivanov(1999)}]{Melosh_1999}
\bibinfo{author}{Melosh, J.}, \bibinfo{author}{Ivanov, B.},
  \bibinfo{year}{1999}.
\newblock \bibinfo{title}{Impact crater collapse}.
\newblock \bibinfo{journal}{Annual Review of Earth and Planetary Sciences}
  \bibinfo{volume}{27}, \bibinfo{pages}{385--415}.
\bibitem[{Michel et~al.(2022)Michel, K{\"u}ppers, Bagatin, Carry, Charnoz,
  de~Leon, Fitzsimmons, Gordo, Green, H{\'{e}}rique, Juzi, Karatekin, Kohout,
  Lazzarin, Murdoch, Okada, Palomba, Pravec, Snodgrass, Tortora, Tsiganis,
  Ulamec, Vincent, W{\"u}nnemann, Zhang, Raducan, Dotto, Chabot, Cheng, Rivkin,
  Barnouin, Ernst, Stickle, Richardson, Thomas, Arakawa, Miyamoto, Nakamura,
  Sugita, Yoshikawa, Abell, Asphaug, Ballouz, Bottke, Lauretta, Walsh, Martino
  and Carnelli}]{Michel_2022}
\bibinfo{author}{Michel, P.}, \bibinfo{author}{K{\"u}ppers, M.},
  \bibinfo{author}{Bagatin, A.C.}, \bibinfo{author}{Carry, B.},
  \bibinfo{author}{Charnoz, S.}, \bibinfo{author}{de~Leon, J.},
  \bibinfo{author}{Fitzsimmons, A.}, \bibinfo{author}{Gordo, P.},
  \bibinfo{author}{Green, S.F.}, \bibinfo{author}{H{\'{e}}rique, A.},
  \bibinfo{author}{Juzi, M.}, \bibinfo{author}{Karatekin, {\"O}.},
  \bibinfo{author}{Kohout, T.}, \bibinfo{author}{Lazzarin, M.},
  \bibinfo{author}{Murdoch, N.}, \bibinfo{author}{Okada, T.},
  \bibinfo{author}{Palomba, E.}, \bibinfo{author}{Pravec, P.},
  \bibinfo{author}{Snodgrass, C.}, \bibinfo{author}{Tortora, P.},
  \bibinfo{author}{Tsiganis, K.}, \bibinfo{author}{Ulamec, S.},
  \bibinfo{author}{Vincent, J.B.}, \bibinfo{author}{W{\"u}nnemann, K.},
  \bibinfo{author}{Zhang, Y.}, \bibinfo{author}{Raducan, S.D.},
  \bibinfo{author}{Dotto, E.}, \bibinfo{author}{Chabot, N.},
  \bibinfo{author}{Cheng, A.F.}, \bibinfo{author}{Rivkin, A.},
  \bibinfo{author}{Barnouin, O.}, \bibinfo{author}{Ernst, C.},
  \bibinfo{author}{Stickle, A.}, \bibinfo{author}{Richardson, D.C.},
  \bibinfo{author}{Thomas, C.}, \bibinfo{author}{Arakawa, M.},
  \bibinfo{author}{Miyamoto, H.}, \bibinfo{author}{Nakamura, A.},
  \bibinfo{author}{Sugita, S.}, \bibinfo{author}{Yoshikawa, M.},
  \bibinfo{author}{Abell, P.}, \bibinfo{author}{Asphaug, E.},
  \bibinfo{author}{Ballouz, R.L.}, \bibinfo{author}{Bottke, W.F.},
  \bibinfo{author}{Lauretta, D.S.}, \bibinfo{author}{Walsh, K.J.},
  \bibinfo{author}{Martino, P.}, \bibinfo{author}{Carnelli, I.},
  \bibinfo{year}{2022}.
\newblock \bibinfo{title}{The {ESA} {Hera} {Mission}: Detailed characterization
  of the {DART} impact outcome and of the binary asteroid (65803) {Didymos}}.
\newblock \bibinfo{journal}{The Planetary Science Journal} \bibinfo{volume}{3},
  \bibinfo{pages}{160}.
\newblock \URLprefix \url{https://doi.org/10.3847%2Fpsj%2Fac6f52},
  \DOIprefix\doi{10.3847/psj/ac6f52}.
\bibitem[{Miklavcic et~al.(2022)Miklavcic, Askari, Sanchez, Quillen and
  Wright}]{Miklavcic_2022}
\bibinfo{author}{Miklavcic, P.M.}, \bibinfo{author}{Askari, H.},
  \bibinfo{author}{Sanchez, P.}, \bibinfo{author}{Quillen, A.C.},
  \bibinfo{author}{Wright, E.}, \bibinfo{year}{2022}.
\newblock \bibinfo{title}{Subsurface dynamics in oblique granular impacts}.
\newblock \bibinfo{journal}{Icarus} \bibinfo{volume}{385},
  \bibinfo{pages}{115089}.
\bibitem[{Naidu et~al.(2020)Naidu, Benner, Brozovic, Nolan, Ostro, Margot,
  Giorgini, Hirabayashi, Scheeres, Pravec, Scheirich, Magri and
  Jao}]{Naidu_2020}
\bibinfo{author}{Naidu, S.}, \bibinfo{author}{Benner, L.},
  \bibinfo{author}{Brozovic, M.}, \bibinfo{author}{Nolan, M.},
  \bibinfo{author}{Ostro, S.}, \bibinfo{author}{Margot, J.},
  \bibinfo{author}{Giorgini, J.}, \bibinfo{author}{Hirabayashi, T.},
  \bibinfo{author}{Scheeres, D.}, \bibinfo{author}{Pravec, P.},
  \bibinfo{author}{Scheirich, P.}, \bibinfo{author}{Magri, C.},
  \bibinfo{author}{Jao, J.}, \bibinfo{year}{2020}.
\newblock \bibinfo{title}{Radar observations and a physical model of binary
  near-earth asteroid {65803} {Didymos}, target of the {DART} mission}.
\newblock \bibinfo{journal}{Icarus} \bibinfo{volume}{348},
  \bibinfo{pages}{113777}.
\newblock \DOIprefix\doi{https://doi.org/10.1016/j.icarus.2020.113777}.
\bibitem[{Poelchau et~al.(2009)Poelchau, Kenkmann and Kring}]{Poelchau_2009}
\bibinfo{author}{Poelchau, M.H.}, \bibinfo{author}{Kenkmann, T.},
  \bibinfo{author}{Kring, D.A.}, \bibinfo{year}{2009}.
\newblock \bibinfo{title}{Rim uplift and crater shape in {Meteor Crater}:
  Effects of target heterogeneities and trajectory obliquity}.
\newblock \bibinfo{journal}{Journal Geophysical Research (Planets)}
  \bibinfo{volume}{114}, \bibinfo{pages}{E01006}.
\bibitem[{Quillen et~al.(2022)Quillen, Neiderbach, Suo, South, Wright,
  Skerrett, S\'anchez, nez, Miklavcic and Askari}]{Quillen_2022}
\bibinfo{author}{Quillen, A.C.}, \bibinfo{author}{Neiderbach, M.},
  \bibinfo{author}{Suo, B.}, \bibinfo{author}{South, J.},
  \bibinfo{author}{Wright, E.}, \bibinfo{author}{Skerrett, N.},
  \bibinfo{author}{S\'anchez, P.}, \bibinfo{author}{nez, F.D.C.},
  \bibinfo{author}{Miklavcic, P.}, \bibinfo{author}{Askari, H.},
  \bibinfo{year}{2022}.
\newblock \bibinfo{title}{Propagation and attenuation of pulses driven by low
  velocity normal impacts in granular media}.
\newblock \bibinfo{journal}{Icarus} \bibinfo{volume}{386},
  \bibinfo{pages}{115139}.
\bibitem[{Rivkin et~al.(2021)Rivkin, Chabot, Stickle, Thomas, Richardson,
  Barnouin, Fahnestock, Ernst, Cheng, Chesley, Naidu, Statler, Barbee, Agrusa,
  Moskovitz, Daly, Pravec, Scheirich, Dotto, Corte, Michel, K{\"u}ppers,
  Atchison and Hirabayashi}]{Rivkin_2021}
\bibinfo{author}{Rivkin, A.S.}, \bibinfo{author}{Chabot, N.L.},
  \bibinfo{author}{Stickle, A.M.}, \bibinfo{author}{Thomas, C.A.},
  \bibinfo{author}{Richardson, D.C.}, \bibinfo{author}{Barnouin, O.},
  \bibinfo{author}{Fahnestock, E.G.}, \bibinfo{author}{Ernst, C.M.},
  \bibinfo{author}{Cheng, A.F.}, \bibinfo{author}{Chesley, S.},
  \bibinfo{author}{Naidu, S.}, \bibinfo{author}{Statler, T.S.},
  \bibinfo{author}{Barbee, B.}, \bibinfo{author}{Agrusa, H.},
  \bibinfo{author}{Moskovitz, N.}, \bibinfo{author}{Daly, R.T.},
  \bibinfo{author}{Pravec, P.}, \bibinfo{author}{Scheirich, P.},
  \bibinfo{author}{Dotto, E.}, \bibinfo{author}{Corte, V.D.},
  \bibinfo{author}{Michel, P.}, \bibinfo{author}{K{\"u}ppers, M.},
  \bibinfo{author}{Atchison, J.}, \bibinfo{author}{Hirabayashi, M.},
  \bibinfo{year}{2021}.
\newblock \bibinfo{title}{The double asteroid redirection test ({DART}):
  Planetary defense investigations and requirements}.
\newblock \bibinfo{journal}{The Planetary Science Journal} \bibinfo{volume}{2},
  \bibinfo{pages}{173}.
\newblock \DOIprefix\doi{10.3847/psj/ac063e}.
\bibitem[{Roddy(1987)}]{Roddy_1978}
\bibinfo{author}{Roddy, D.J.}, \bibinfo{year}{1987}.
\newblock \bibinfo{title}{Pre-impact geologic conditions, physical properties,
  energy calculations, meteorite and initial crater dimensions and orientations
  of joints, faults and walls at {Meteor Crater}, {Arizona}}, in:
  \bibinfo{booktitle}{Proc. Lunar Planet. Sci. Conf., 9th},
  \bibinfo{organization}{Lunar and Planetary Institute and American Geophysical
  Union}. \bibinfo{publisher}{New York: Pergamon Press}. pp.
  \bibinfo{pages}{3891--3930}.
\bibitem[{Rosas and Lindenberg(2018)}]{Rosas_2018}
\bibinfo{author}{Rosas, A.}, \bibinfo{author}{Lindenberg, K.},
  \bibinfo{year}{2018}.
\newblock \bibinfo{title}{Pulse propagation in granular chains}.
\newblock \bibinfo{journal}{Physics Reports} \bibinfo{volume}{735},
  \bibinfo{pages}{1--37}.
\bibitem[{Schmidt and Housen(1987)}]{Schmidt_1987}
\bibinfo{author}{Schmidt, R.M.}, \bibinfo{author}{Housen, K.R.},
  \bibinfo{year}{1987}.
\newblock \bibinfo{title}{Some recent advances in the scaling of impact and
  explosion cratering}.
\newblock \bibinfo{journal}{International Journal of Impact Engineering}
  \bibinfo{volume}{5}, \bibinfo{pages}{543--560}.
\bibitem[{Sharpton(2014)}]{Sharpton_2014}
\bibinfo{author}{Sharpton, V.L.}, \bibinfo{year}{2014}.
\newblock \bibinfo{title}{Outcrops on lunar crater rims: Implications for rim
  construction mechanisms, ejecta volumes and excavation depths}.
\newblock \bibinfo{journal}{J. Geophys. Res. Planets} \bibinfo{volume}{119},
  \bibinfo{pages}{154--168}.
\bibitem[{Somfai et~al.(2005)Somfai, Roux, Snoeijer, van Hecke and van
  Saarloos}]{Somfai_2005}
\bibinfo{author}{Somfai, E.}, \bibinfo{author}{Roux, J.N.},
  \bibinfo{author}{Snoeijer, J.H.}, \bibinfo{author}{van Hecke, M.},
  \bibinfo{author}{van Saarloos, W.}, \bibinfo{year}{2005}.
\newblock \bibinfo{title}{Elastic wave propagation in confined granular
  systems}.
\newblock \bibinfo{journal}{Phys. Rev. E} \bibinfo{volume}{72}.
\bibitem[{Sturm et~al.(2016)Sturm, Kenkmann and Hergarten}]{Sturm_2016}
\bibinfo{author}{Sturm, S.}, \bibinfo{author}{Kenkmann, T.},
  \bibinfo{author}{Hergarten, S.}, \bibinfo{year}{2016}.
\newblock \bibinfo{title}{Ejecta thickness and structural rim uplift
  measurements of {Martian} impact craters: implications for the rim formation
  of complex impact craters}.
\newblock \bibinfo{journal}{J. Geophys. Res.} \bibinfo{volume}{121},
  \bibinfo{pages}{1026--1053}.
\bibitem[{Tell et~al.(2020)Tell, Drei{\ss}igacker, Tchapnda, Yu and
  Sperl}]{Tell_2020}
\bibinfo{author}{Tell, K.}, \bibinfo{author}{Drei{\ss}igacker, C.},
  \bibinfo{author}{Tchapnda, A.C.}, \bibinfo{author}{Yu, P.},
  \bibinfo{author}{Sperl, M.}, \bibinfo{year}{2020}.
\newblock \bibinfo{title}{Acoustic waves in granular packings at low
  confinement pressure}.
\newblock \bibinfo{journal}{Review of Scientific Instruments}
  \bibinfo{volume}{91}, \bibinfo{pages}{033906}.
\bibitem[{Thomas and Robinson(2005)}]{thomas05}
\bibinfo{author}{Thomas, P.C.}, \bibinfo{author}{Robinson, M.S.},
  \bibinfo{year}{2005}.
\newblock \bibinfo{title}{Seismic resurfacing by a single impact on the
  asteroid 433 {Eros}}.
\newblock \bibinfo{journal}{Nature} \bibinfo{volume}{436},
  \bibinfo{pages}{366--369}.
\bibitem[{Thomsen et~al.(1980)Thomsen, Austin, Ruhl, Orphal and
  Schultz}]{Thomsen_1980}
\bibinfo{author}{Thomsen, J.M.}, \bibinfo{author}{Austin, M.G.},
  \bibinfo{author}{Ruhl, S.F.}, \bibinfo{author}{Orphal, D.L.},
  \bibinfo{author}{Schultz, P.H.}, \bibinfo{year}{1980}.
\newblock \bibinfo{title}{The detailed application of {Maxwell's} {Z}-model to
  laboratory-scale impact cratering calculations,}, in:
  \bibinfo{editor}{Merrill, R.B.}, \bibinfo{editor}{Schultz, P.H.} (Eds.),
  \bibinfo{booktitle}{[Abstracts of] Papers Presented to the Conference on
  Multi-ring Basins: Formation and Evolution}, p.~\bibinfo{pages}{92}.
\bibitem[{Tzeremes et~al.(2019)Tzeremes, Jones, Hernandez, Sousa, Pollini,
  Pache, Haesler and Carnelli}]{altimeter}
\bibinfo{author}{Tzeremes, G.D.}, \bibinfo{author}{Jones, D.},
  \bibinfo{author}{Hernandez, M.}, \bibinfo{author}{Sousa, T.},
  \bibinfo{author}{Pollini, A.}, \bibinfo{author}{Pache, C.},
  \bibinfo{author}{Haesler, J.}, \bibinfo{author}{Carnelli, I.},
  \bibinfo{year}{2019}.
\newblock \bibinfo{title}{Altimetry, imaging and landing location selection
  lidars for esa space applications}, in: \bibinfo{booktitle}{IGARSS 2019 -
  2019 IEEE International Geoscience and Remote Sensing Symposium}, pp.
  \bibinfo{pages}{4775--4778}.
\newblock \DOIprefix\doi{10.1109/IGARSS.2019.8900519}.
\bibitem[{Wright et~al.(2022)Wright, Quillen, Sanchez, Schwartz, Nakajima,
  Askari and Miklavcic}]{Wright_2022}
\bibinfo{author}{Wright, E.}, \bibinfo{author}{Quillen, A.C.},
  \bibinfo{author}{Sanchez, P.}, \bibinfo{author}{Schwartz, S.R.},
  \bibinfo{author}{Nakajima, M.}, \bibinfo{author}{Askari, H.},
  \bibinfo{author}{Miklavcic, P.}, \bibinfo{year}{2022}.
\newblock \bibinfo{title}{Ricochets on asteroids {II}: Sensitivity of
  laboratory experiments of low velocity grazing impacts on substrate grain
  size}.
\newblock \bibinfo{journal}{Icarus, accepted for publication}
  \bibinfo{volume}{376}, \bibinfo{pages}{114868}.
\bibitem[{Wright et~al.(2020)Wright, Quillen, South, Nelson, S{\'{a}}nchez,
  Siu, Askari, Nakajima and Schwartz}]{Wright_2020b}
\bibinfo{author}{Wright, E.}, \bibinfo{author}{Quillen, A.C.},
  \bibinfo{author}{South, J.}, \bibinfo{author}{Nelson, R.C.},
  \bibinfo{author}{S{\'{a}}nchez, P.}, \bibinfo{author}{Siu, J.},
  \bibinfo{author}{Askari, H.}, \bibinfo{author}{Nakajima, M.},
  \bibinfo{author}{Schwartz, S.R.}, \bibinfo{year}{2020}.
\newblock \bibinfo{title}{Ricochets on asteroids: Experimental study of low
  velocity grazing impacts into granular media}.
\newblock \bibinfo{journal}{Icarus} \bibinfo{volume}{351},
  \bibinfo{pages}{113963}.
\newblock \DOIprefix\doi{10.1016/j.icarus.2020.113963}.
\bibitem[{Yamamoto et~al.(2009)Yamamoto, Barnouin-Jha, Toriumi, Sugita and
  Matsui}]{Yamamoto_2009}
\bibinfo{author}{Yamamoto, S.}, \bibinfo{author}{Barnouin-Jha, O.S.},
  \bibinfo{author}{Toriumi, T.}, \bibinfo{author}{Sugita, S.},
  \bibinfo{author}{Matsui, T.}, \bibinfo{year}{2009}.
\newblock \bibinfo{title}{An empirical model for transient crater growth in
  granular targets based on direct observations}.
\newblock \bibinfo{journal}{Icarus} \bibinfo{volume}{203},
  \bibinfo{pages}{310--319}.
\bibitem[{Yasui et~al.(2015)Yasui, Matsumoto and Arakawa}]{yasui15}
\bibinfo{author}{Yasui, M.}, \bibinfo{author}{Matsumoto, E.},
  \bibinfo{author}{Arakawa, M.}, \bibinfo{year}{2015}.
\newblock \bibinfo{title}{Experimental study on impact-induced seismic wave
  propagation through granular materials}.
\newblock \bibinfo{journal}{Icarus} \bibinfo{volume}{260},
  \bibinfo{pages}{320--331}.
\bibitem[{Zhai et~al.(2020)Zhai, Herbold and Hurley}]{Zhai_2020}
\bibinfo{author}{Zhai, C.}, \bibinfo{author}{Herbold, E.B.},
  \bibinfo{author}{Hurley, R.C.}, \bibinfo{year}{2020}.
\newblock \bibinfo{title}{The influence of packing structure and interparticle
  forces on ultrasound transmission in granular media}.
\newblock \bibinfo{journal}{Proceedings of the National Academy of Sciences}
  \bibinfo{volume}{117}, \bibinfo{pages}{16234--16242}.

\end{thebibliography}

\if \ispreprint1
\begin{table*}[]
    \centering
        \caption{Nomenclature}
    \begin{tabular}{lll}
    \hline
      Final displacements   &  $\delta_{r,f}$ $\delta_{r,f}$\\
      Peak displacements & $\delta_{r,pk}$ $\delta_{z,pk}$ \\
      Peak velocities  & $v_{r,pk}$, $v_{z,pk}$ \\
      Impact velocity & $v_{imp}$ \\
      Projectile mass & $M_p$ \\
      Projectile radius & $R_p$  \\
      Substrate bulk density & $\rho_s$\\
      Projectile density & $\rho_p$ \\
      Pulse propagation velocity & $v_P$ \\
      Distance from site of impact & $r$ \\
      Crater radius & $R_c$\\
      Crater depth & $h_c$ \\
      Cylindrical coordinates & $(R,\phi,z)$ \\
      Dimensionless momentum scaling parameter & $B_{\rm eff}$ \\
      Spherical coordinates & $(r,\vartheta,\phi)$ \\
      Ratio of displacement to crater radius & $\mathcal{K}_r = \delta_{r,f}(R_c)/R_c$ \\
      Ratio of uplift to crater radius & $\mathcal{K}_z = \delta_{z,f}(R_c)/R_c$ \\
      Dimensionless scaling parameters & $\pi_2, \pi_4$ \\
      Gravitational acceleration (Earth) & $g$ \\
      Gravitational acceleration  & $g_{\rm eff}$ \\
      Crater formation time & $\tau_{ex} = \sqrt{R_c/g}$ \\
      Vertical coordinate & $\bm \hat z$ \\
      Structural uplift crater rim height & $h_u$ \\
         \hline
    \end{tabular}
    \label{tab:my_label}
\end{table*}
\fi

\end{document}